\documentstyle[psfig]{mn}

\def\aq{\mbox{$a_4$}}

\def\rms{\mbox{\rm r.m.s.}}

\def\muem{\mbox{$\langle\mu\rangle_{\rm e}$}}
\def\muen{\mbox{$\langle\mu\rangle_{n}$}}
\def\mga{\mbox{mag arcsec$^{-2}$}}
\def\sbu{\mbox{$B$ mag arcsec$^{-2}$}}
\def\re{\mbox{$r_{\rm e}$}}

\def\CCD{\mbox{C$^2$D}}
\def\kms{\mbox{\,km~s$^{-1}$}}

\def\Iem{\mbox{$\langle I \rangle_{\rm e}$}}

\def\Dmu{\mbox{$\Delta\mu$}}
\def\Dmufv{\mbox{$\Delta\mu_{\rm FV}$}}
\def\Dmucv{\mbox{$\Delta\mu_{\rm CV}$}}
\def\Dldn{\mbox{$\Delta\log(D_n)$}}
\def\Dlre{\mbox{$\Delta\log(r_{\rm e})$}}
\def\Dldnfv{\mbox{$\Delta\log(D_n^{\rm FV})$}}
\def\Dldncv{\mbox{$\Delta\log(D_n^{\rm CV})$}}
\def\Dlrefv{\mbox{$\Delta\log(r_{\rm e}^{\rm FV})$}}
\def\Dlrecv{\mbox{$\Delta\log(r_{\rm e}^{\rm CV})$}}
\def\rq{\mbox{r$^{1/4}$}}
\def\rn{\mbox{r$^{1/m}$}}

\def\dn{\mbox{$D_n$}}
\def\dnae{\mbox{$D_n/2r_{\rm e}$}}
\def\dns{\mbox{$D_n$--$\sigma$}}
\def\dnsmu{\mbox{$D_n$--$\sigma$--$\Delta\mu$}}
\def\dnaemue{\mbox{$(D_n/A_{\rm e})$--$\muem$}}
\def\lsmu{\mbox{$L$--$\sigma$--$\mu$}}
\def\lrem{\mbox{$\log(m)$--$\log(\re)$}}
\def\FJ{\mbox{FJ}}

\def\etal{{\rm et al.\/}}
\def\cf{{\rm cf.\/}}
\def\ie{{\rm i.e.\/}}
\def\eg{{\rm e.g.\/}}
\def\ph{\phantom}
\def\IRTF{IR-TF}

\def\GV{de\,Vau\-cou\-leurs}

\newcommand{\mAJ}[3]{#1, AJ, {#2}, #3}
\newcommand{\mApJ}[3]{#1, ApJ, {#2}, #3}
\newcommand{\mApJS}[3]{#1, ApJS, {#2}, #3}
\newcommand{\mAeA}[3]{#1, A\&A, {#2}, #3}

\newcommand{\mAeAS}[3]{#1, A\&AS, {#2}, #3}

\newcommand{\mMN}[3]{#1, MNRAS, {#2}, #3}

\newcommand{\mNat}[3]{#1, Nature, {#2}, #3}
\newcommand{\mPASP}[3]{#1, PASP, {#2}, #3}

\title[Relative distances to the Virgo, Fornax, and Coma clusters] {The
relative distances to the Virgo, Fornax, and Coma clusters of galaxies
through the \dns\ and the Fundamental Plane relations}
\author[M. D'Onofrio et al.]{Mauro D'Onofrio$^1$, Massimo
Capaccioli$^{2,3}$, Simone R. Zaggia$^{1,3}$, and Nicola Caon$^4$\\
$^1$Dipartimento di Astronomia, Universit\`a di Padova, 
vicolo dell'Osservatorio 5, I-35122 -- Padova -- Italy \\
$^2$Dipartimento di Scienze Fisiche, Universit\`a di Napoli,
Mostra d'Oltremare, Padiglione 19, I-80125 -- Napoli -- Italy \\
$^3$Osservatorio Astronomico di Capodimonte, 
salita Moiariello 16, I-80131 -- Napoli -- Italy \\ 
$^4$Space Telescope Science Institute, 3700 S. Martin Drive -- Baltimore, MD 21218 -- USA --- Affiliated to the \\
\ Astrophysics Division, Space Science Department, ESA}

\date{Received ...; accepted 24 march 1997  \hfill {\it To G\`erard de Vaucouleurs}}

\begin{document}

\label{firstpage}

\maketitle

\begin{abstract}
We derive the relative distances to the Virgo, Fornax, and Coma clusters
of galaxies by applying the \dns\ and the Fundamental Plane (FP) relations
to the data of the homogeneous samples of early--type galaxies studied by Caon
\etal\ (1990, 1994), Lucey \etal\ (1991a,b), and J{\o}rgensen \etal\
(1992, 1995a,b). 
The two distance indicators give consistent results, the relative distance
moduli to Fornax and Coma with respect to Virgo being
$\Dmufv = (0.45\pm0.15)$ mag and $\Dmucv = (3.55\pm0.15)$ mag
respectively. The formal error on $\Dmu$ may be as small as 0.07 mag
($\sim3\%$ in distance), provided that all the sources of bias are taken
into account and a correct statistical approach is used. 
Unfortunately, much of the actual uncertainty
in the relative distance of the clusters ($\sim 12-15\%$), is due to the
existence of systematic departures in the measurements of the velocity 
dispersions among the various datasets, and to the corrections for
aperture effects.

The above result for the Fornax cluster is supported by the \lsmu\ relation
and, with lesser accuracy, by the \lrem\ relations.
Our value of $\Dmufv$ is in fair agreement with the one derived using
planetary nebulae and SNe--Ia, while is in open contrast with that coming
from surface brightness fluctuations, globular clusters luminosity
function, and infrared Tully-Fisher relation. In our data Coma appears
slightly nearer than indicated by the other distance indicators, but now a
better agreement with the Tully-Fisher relation seems to exist.

We show that for the galaxies of the Virgo and Fornax clusters
the residuals of the \dns\ relation do not correlate with the
effective surface brightness \muem. There is also no correlation of the
residuals of the \dns\ and FP relations with the total luminosity of the
galaxies, with the ellipticity and with the isophotal shape parameter \aq.
Instead, a correlation seems to exist with the maximum rotation velocity of
the galaxies, $V_{\rm m}$, with the $(V/\sigma)$ ratio, and with the exponent
$m$ of the \rn\ fit to the major axis light profiles of the galaxies. If
confirmed, these effects introduce a systematic bias in both relations when
used as distance indicators.
\end{abstract}

%

\section{Introduction}\label{sec1}
Empirical scaling laws built using the main structural parameters of
galaxies such as characteristic radii, mean surface brightnesses, and
central velocity dispersions, have become quite fashionable in the last
decade. In particular, the relations holding for early-type galaxies are
believed to be useful tools for the understanding of the structure,
formation, and evolution of the kinematically hot stellar systems (hereafter 
early-type galaxies), and to be also quite accurate distance indicators
(DIs) for these objects.

One of such relations is the so-called Fundamental Plane (hereafter FP;
Djorgovski \& Davis 1987), which, in its original formulation, binds the
central velocity dispersion, $\sigma_0$, and the average surface brightness
within the effective isophote (that encircling half of the total light of a
galaxy in a given photometric band), \muem, to the mean radius of this same
isophote, named effective radius $r_{\rm e}$: $\log(r_{\rm e}) =
a\,\log(\sigma_0) + b\langle\mu\rangle_{\rm e} + c$.

A second scaling law is the \dns\ relation (Dressler \etal\ 1987):
$\dn\propto\sigma_0^\gamma$, where \dn\ is the diameter of the circular
aperture within which the average corrected surface brightness equals a
fixed value \muen\, usually 20.75 \sbu, and $\gamma$ is a constant of the 
order of $3/2$. To a first order approximation the \dns\
relation is an edge-on view of the FP (Dressler \etal\ 1987).
In this case the surface brightness parameter is hidden in the 
definition of \dn.
 
The slope and the small thickness measured for the FP and the \dns\ relations
imply a narrow range of density structure, velocity anisotropy, and mean
age among elliptical galaxies, which are excellent properties for DIs. 
Actually these relations have provided much of the present information on the
large scale flows in the local universe (Lynden-Bell \etal\ 1988). 

Two complementary strategies have been developed. On the one hand the
enormous progress in the observational techniques and data reduction
procedures has produced larger and larger samples and increasingly better
distance determinations of galaxies. On the other hand much effort has been
put in the last years in quantifying internal errors, biases, calibration
uncertainties, second parameter corrections, environmental influences, and
behaviour of the residuals in both relations; see \eg\ Lucey \etal\
(1991a; b = LGCT), Jacoby \etal\ (1992), J{\o}rgensen \etal\ (1992; =
JFK, 1995a,b), Davies \etal\ (1993), Saglia \etal\ (1993), van Albada \etal\
(1993). In conclusion, from the original 32\% uncertainty given by the
Faber \& Jackson (1976; \FJ) relation, the error has been progressively
reduced by the coming of the \dns\ and of the FP relations, being
nowadays as low as 10\% in the relative distances between clusters (Jacoby
\etal\ 1992).

Despite this success, a number of problems are still open. Which of the
two relations, \dns\ and FP, is more suiteable as DI\,? Or, in other words,
which of the two is less affected by observational errors and biases\,?
Can S0 galaxies be included in these relations\,? Many papers have
addressed the above questions with contradictory results.

JFK state that the FP should be preferred, because the \dns\ relation has
a larger scatter and suffers from a surface brightness bias. Van Albada
\etal\ (1993) find that the \dns\ relation has the same accuracy of the FP,
provided that a proper surface brightness correction is applied. Lucey
\etal\ (1991a) and LGCT derive contradictory evidences for such a bias in
their samples, and Saglia \etal\ (1993) show that the scatter around both
relations can be reduced by rejecting E galaxies with inner disk
components. Finally, Davies \etal\ (1993) point out that only galaxies in
a restricted range of surface brightness should be used in the \dns\
relation in order to avoid contaminations.

We argue here that the \dns\ and the FP relations have substantially the same
accuracy -- at least as long as the distance to nearby clusters is
concerned --  
by applying them to the measure of the relative distances to the Fornax
and Coma clusters with respect to Virgo. The motivation of
this work is twofold. Firstly, the relative distances to these clusters
are far from being accurately known (cf. Section~\ref{sec2}), and
secondly, today we can exploit an accurate and homogeneous data-set of
photometric and spectroscopic measurements for the galaxies of these three
clusters: Caon \etal\ (1990, 1994, hereafter \CCD), D'Onofrio \etal\
(1995), LGCT, JFK, and J{\o}rgensen \etal\ (1995a, b). 
The study of a volume-limited sample of galaxies has
the advantage that one can look at the intrinsic properties of the \dns\
and FP relations by means of objects which are probably coeval, and reduce as
much as possible the principal sources of errors, in particular those related
to the unknown distances of the galaxies. 

The paper is organized as follows. Section~\ref{sec2} reviews the current
determinations of the relative distances to the Virgo, Fornax, and Coma
clusters, as measured by the relevant DIs. Our determinations of the
relative distances to the three clusters using the \dns\ and FP relations
are presented in Sections~\ref{sec3} and \ref{sec4} respectively, together
with a discussion of the errors intrinsic to both methods and a comparison
of the results with other DIs. In Section~\ref{sec5} we analyze the
behaviour of the residuals of the \dns\ and FP relations by exploiting the
advantage of our volume-limited, accurate, and homogeneous (in the
methodological approach and in the data analysis) sample of galaxies, and
we perform some further tests on the \dns\ relation by varying the
definitions of $D_n$ and $\sigma$. In Section~\ref{sec6} we present the
\lrem\ relation used as DI for the Fornax cluster. Conclusions are drawn
in Section~\ref{sec7}.

\section{The relative distances to the Virgo, Fornax, and Coma
clusters}\label{sec2} 
In his review of the extragalactic distance scale, \GV\ (1993) showed 
that the differential distance modulus (\Dmu) between the Fornax and the
Virgo cluster is, on average, $\Dmufv=+0.16$ and $+0.38$ mag for
the {\it long\/} and for the {\it short\/} distance scale respectively. 
In other words, according to this study, both scales place Virgo closer
than Fornax.

The trend is not confirmed by all DIs. Table~1 lists the measurements
of \Dmufv $=\mu_F - \mu_V$, sorted according to the
method used; we have essentially ignored the old measurements, already
revisited by \GV. A quick look to the Table shows that the inner ring
diameters of spirals, the globular cluster luminosity function (GCLF), the
surface brightness fluctuations (SBF), the \lsmu\ relation, the IR
Tully--Fisher relation (\IRTF), and the scale length of dE galaxies, place
the Fornax cluster closer than Virgo, in marked contrast with the
results from the brightest cluster members (BCM), the \dns\ relation, the
planetary nebulae luminosity function (PNLF), and the SNe--Ia.
Taking a plain average, the \Dmufv\ turns into a salomonical
$\sim0.0$, with a \rms\ of $0.24$ mag.

\begin{table*}
\begin{tabular}{cllp{8cm}}
\multicolumn{4}{c}{Table 1: Distance modulus of the Fornax cluster
relative to that of Virgo} \\
\hline\hline\noalign{\smallskip}
$\Dmufv$ & Method & Reference & Notes \\
\noalign{\smallskip}\hline\noalign{\smallskip}
$+0.48\pm0.20$ & BCM lum. & de Vaucouleurs 1977 & Uncertain estimate of
the error\\ 
$-0.37\pm0.30$ & Ring diam. & Buta \& de Vaucouleurs 1983 &
Based on inner rings of 4 spirals in Fornax and 17 in Virgo\\
$+0.15\pm0.28$ & \dns\ & Davies \etal\ 1993 & Calibrated through the Leo
Group set at a distance of 10 Mpc\\ 
$+0.13\pm0.20$ & \dns\ & Dressler \etal\
1987 & The estimated error concerns galaxies in clusters\\ 
$+0.25\pm0.31$
& \dns\ & Faber \etal\ 1989 & See McMillan \etal\ (1993)\\ 
$-0.14\pm0.28$
& \lsmu\ & Pierce 1989 & Assuming Leo Group at 10 Mpc\\ 
$+0.00\pm0.22$ &
$L$--$\Sigma$ & Ferguson \& Sandage 1988 & See McMillan \etal\ (1993)\\
$-0.15\pm0.13$ & SBF & Tonry 1991 & Assuming $\mu=24.43$ mag for M31 and
M32 and no correction for absorption\\ 
$-0.17\pm0.34$ & SBF & Buzzoni 1993
& Assuming M32 at 0.74 Mpc\\ 
$+0.08\pm0.11$ & SBF & Tonry 1991 & See
McMillan \etal\ (1993)\\ 
$-0.50\pm0.20$ & GCLF & Geisler \& Forte 1990 &
Match of the LF of NGC\,1399 and M87\\ 
$-0.11\pm0.14$ & GCLF & Bridges
\etal\ 1991 & Using the LF of NGC\,1399 and a composite Virgo LF\\
$-0.16\pm0.15$ & dE s. length & Bothun \etal\ 1989 & From the scale length
of $\sim25$ dwarf ellipticals in both clusters\\ 
$-0.25\pm0.23$ & \IRTF &
Aaronson \etal\ 1989 & H-band TF for 16 spirals in Virgo and 7 in Fornax \\
$+0.24\pm0.10$ & PNLF & McMillan \etal\ 1993 & From the PNLF of NGC\,1316,
NGC\,1399, and NGC\,1404\\ 
$+0.03\pm0.10$ & PNLF & Ciardullo \etal\ 1991 &
>From de Vaucouleurs (1993)\\ 
$+0.15\pm0.94$ & SNe--Ia & Hamuy \etal\ 1991
& B-band maximum magnitude of SNe 1980N and 1981D in NGC\,1316 and SN 1984A in
NGC\,4419\\ 
$+0.26\pm0.71$ & SNe--Ia & Hamuy \etal\ 1991 & Maximum V-band 
magnitude of the same objects\\ 
$+0.09\pm0.14$ & SNe--Ia & Hamuy \etal\ 1991 &
Maximum H-band magnitude of the same objects\\
\noalign{\smallskip}\hline
\end{tabular}
\end{table*}

A few comments to Table~1 are in order. The SBF method gives distances
systematically lower by 10-30\% than the \dns\ relation, probably due
to the adopted calibration (Sandage \& Tammann 1995). The values of
\Dmu\ obtained from the analysis of the GCLF strongly depend on
the morphological types of the selected galaxies and on metallicity
effects, while those provided by the PNLF method are in good agreement,
within the errors, with the \dns\ relation (McMillan \etal\ 1993). 
For both methods
however, the number of galaxies in common with our \dns\ relation
is quite small, and a conclusive comparison cannot be made. The filamentary
and complex structure of the Virgo cluster probably accounts for the
smaller distance found by the \IRTF\ relation, since the spiral galaxies have a
different spatial distribution than ellipticals. The negative \Dmu\ value
given by Pierce (1989) through the \lsmu\ relation is probably the consequence
of the small size of his sample: we obtain $\Dmufv = 0$ by applying the \dns\
relation to the galaxies of our sample in common with his. This is a clear 
demonstration that a ``representative'' statistical sample of galaxies in
a cluster, selected according to well defined criteria, must be used 
to gauge relative distances through the \dns\ and FP relations (see next
section). 

\begin{table*}
\begin{tabular}{cllp{8cm}}
\multicolumn{4}{c}{Table 2: Distance modulus of the Coma cluster relative
to that of Virgo} \\
\hline\hline\noalign{\smallskip}
$\Dmucv$ & Method & Reference & Notes \\
\hline \\[-7pt]
$+3.72\pm0.30$ & Max. of 5 SNe & de Vaucouleurs 1985 & From Table 16B in
de Vaucouleurs (1993)\\ 
$+3.65\pm0.20$ & \dns\ & Dressler \etal\ 1987 & From a sample of 20 Es in
Virgo and 28 Es in Coma\\
$+3.99\pm0.65$ & $D_n$--$Mg_2$ & Dressler \etal\ 1987 & The error is the
35\% \rms\ uncertainty of the method\\
$+3.82\pm0.10$ & \dns\ & Feigelson \& Babu 1992 & Data from LGCT and
Dressler \etal\ (1987)\\
$+3.75\pm0.20$ & SNe--Ia & Capaccioli \etal\ 1990 & Data from the Asiago
SN Catalogue of Barbon \etal\ (1989)\\
$+3.60\pm0.30$ & SNe--Ia & Capaccioli \etal\ 1990 & Using SNe--Ia in 4 E
galaxies\\
$+3.69\pm0.12$ & \IRTF & Aaronson \& Mould 1986 & See Table 4 of
Capaccioli \etal\ (1990)\\
$+3.66\pm0.14$ & $U-B$ of E & Sandage 1972 & Colour-magnitude (CM) diagram.
Revised to 3.5$\pm0.2$ by Aaronson \etal\ (1981)\\
$+3.70\pm0.17$ & B and H TF & Giraud 1986 & Hybrid form of the TF
relation\\
$+3.76\pm0.12$ & $L$--$\sigma$ & Lucey 1986 & Through fixed metric
apertures\\ 
$+3.75\pm0.18$ & $V_{26}$--$\sigma$ & Dressler 1984 & $V_{26}$ from Sandage \&
Visvanathan (1978) \\
$+4.07\pm0.75$ & Reduced radii & Gudehus 1991 & Mean error for
a bright isolated galaxy\\
$+3.42\pm0.22$ & B--band TF & Rood \& Williams 1992 & For M31--like
galaxies\\
$+3.70\pm0.09$ & Mean of 3 & Bower \etal\ 1992 & Mean of CM diagram,
colour-diameter and \FJ\ relations \\
$+3.59\pm0.06$ & CM diagr. for E & Bower \etal\ 1992 & As above for E
galaxies only\\ $+3.80\pm0.06$ & Mult. analysis & Vader 1986 & Comparison
between the mass-luminosity relations\\
$+3.60\pm0.17$ & IR CM diagram & Christensen 1991 & Unbiased value for
spiral galaxies\\
$+3.90\pm0.10$ & Mean of 9 & Tammann \& Sandage 1985& Mean of 9
determinations from the literature\\ $+4.18\pm0.19$ & Mean of 3 & Sandage
\& Tammann 1982& See Table 4 of Capaccioli \etal\ (1990)\\
$+3.60\pm0.40$ & B--band TF & Fukugita \etal\ 1993 & Virgo centre placed
at 15 Mpc\\
\noalign{\smallskip}\hline
\end{tabular}
\end{table*}

In a similar way Table~2 reviews the determinations of the relative
distance between the Virgo and the Coma clusters. The average differential
distance mudulus is now
$3.75\pm0.18$ mag, but the individual determinations of \Dmucv\ span over
an interval of $\sim0.6$ mag. Most of the measurements range however
between 3.60 and 3.80 mag. If the centre of the Virgo cluster is taken at
18.3 Mpc (Capaccioli \etal\ 1990), considering the broad range of values
in Table~2, the distance to Coma ranges from 88.4 and
119.3 Mpc.

We note that the \dns\ relation of Dressler \etal\ (1987) gives results in good
agreement with the SNe--Ia method and with the TF determinations. It is
also apparent from the above Tables that the principal DIs provide quite
different distances for each cluster. We will not try to give here an
explanation for such discrepancies, since this would require a large
number of distance determinations in common between the various DIs. In
the following we will address only the problems encountered with the \dns\
and FP relations, when they are used as DIs.

\section{The relative distances to the Virgo, Fornax, and Coma clusters
through the \dns\ and the FP relations}\label{sec3}
There are various sources of errors affecting the determinations of the
relative distances to nearby galaxy clusters with the \dns\ and FP
relations. The selection criteria of the galaxy sample play an important
part in this game, together with the accuracy and homogeneity of the
photometric and spectroscopic data, the definition of the variables
involved in the relations, and the statistical regression analysis that is
used.

A number of questions still awaits for a clarification. Does it exist a
``representative'' sample of early-type galaxies in clusters providing
unbiased distance determinations\,? Can S0 galaxies be included in the \dns\
and FP relations\,? Which are the ``best'' parameters to use\,? For instance,
shall we use the effective radius \re\ derived from a model--dependent fit to
the light profiles (e.g. \rq\ law), rather than the radius of the isophote
which encircles half the total luminosity, whatever the light profile be\,?
Shall we choose the central velocity dispersion $\sigma_0$ or an average
within a given radial range\,? Is it better suited the diameter \dn\ at
$\mu_B=20.75$ \mga or the diameter relative to a different mean surface
brightness level\,? We have no clear answers to any of such questions yet. 

Unfortunately, the accuracy and homogeneity of the spectroscopic data
is still low compared with those of photometric data. Extended rotation curves
and velocity dispersion profiles do exist for a few galaxies only
(usually the brightest ones), and in most cases the full velocity field 
of the objects is poorly known.
Systematic and/or random differences in the velocity dispersion measurements 
are easily found in the literature for the same objects and 
likely arise from
the use of different methods of data reduction (Fourier Quotient, Cross
Correlation, Fourier Correlation Quotient, etc.) and to the corrections for
the aperture size effects (J{\o}rgensen \etal\ 1995b). 

Presently the Virgo, Fornax, and Coma clusters offer the best studied
galaxy samples for addressing the above questions, since the
available data approximately satisfy the required criteria of homogeneity
and accuracy needed for the analysis of these problems.

\subsection{The samples}\label{sec3a}
The \CCD\ set of B-band photometric parameters for Virgo E and
non-barred S0 galaxies is 80\% complete down to $B_T=14$ mag (according to
the catalog of Binggeli \etal, 1985, hereafter BST), and that of Fornax is
100\% complete down to $B_T=15$ mag (according to the membership list of
Ferguson, 1989). Assuming the same distance of 18.3 Mpc to both
clusters, the samples are respectively complete (and volume-limited) down
to $M_B =-17.3$ and $M_B =-16.3$ mag.

The \dn\ diameters, effective radii \re, and effective mean surface
brightnesses \muem, are given by \CCD\ with an accuracy of about
5\%, 20\%, and $\sim0.3$ mag ($\sim 30\%$) respectively. 
The above parameters have not been corrected for seeing, galactic absorption, 
internal extinction and redshift.
All these corrections are very small for the galaxies in
both clusters, as explained in \CCD.

For the photometric parameters of the Coma galaxies we made use of the
data of LGCT, JFK, and J{\o}rgensen \etal\ (1995a), which have a similar degree
of homogeneity and accuracy as \CCD. These data have been
corrected by the authors for the above cited effects.

\begin{figure} 
\psfig{figure=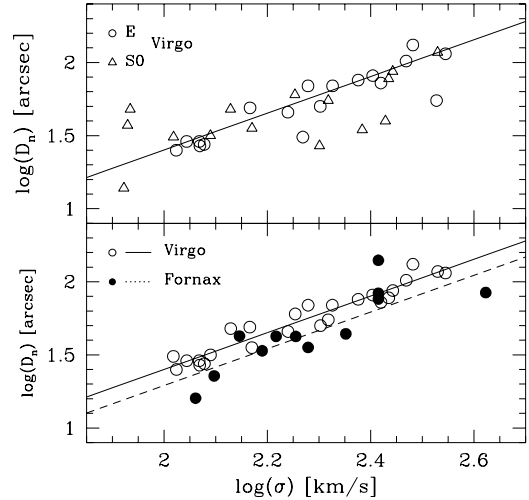,width=8cm,angle=0}
\caption[]{{\it Panel a)} Distribution of the kinematical subset of the
Virgo galaxies of the \CCD\
sample in the $\log\dn$--$\log\sigma$ plane. The solid line gives the fit
to the ``best sample'', obtained by rejecting the objetcs
deviating largely from the mean relation.
{\it Panel b)} The same diagram for the Virgo (``best sample'') and Fornax
early-type galaxies. The dashed line is the fit of minimum scatter for the
galaxies of the Fornax cluster, adopting the same slope of the Virgo
relation.}
\label{fig1}
\end{figure}
 
Unfortunately, the spectroscopic data are far less homogeneous. We have
measured the rotation curves $V(r)$ and velocity dispersion profiles
$\sigma(r)$ of 15 galaxies of the \CCD\ Fornax sample (D'Onofrio \etal\ 1995),
while for the Virgo galaxies we collected central velocity
dispersion measurements from the literature: Whitmore \etal\ (1985, WMT),
Dalle Ore \etal\ (1991, DFJS), Davoust \etal\ (1985, DPV), Faber \etal\
(1989, 7-Sam), McElroy (1995). 
For the Virgo cluster we were able to find kinematical data for
12 galaxies only (10 E and 2 S0): 
NGC\,4374, 4486, 4636 (Davies \& Birkinshaw 1988),
NGC\,4365, 4621 (Bender 1988a,b), NGC\,4387, 4478 (Davies \etal\ 1983), 
NGC\,4406, 4472, 4649 (Franx \etal\ 1989), NGC\,4459 (Peterson 1978) 
and NGC\,4473 (Young \etal\ 1978). 

For the galaxies of the Coma cluster we used the central velocity dispersions
quoted by LGCT, JFK (who homogeneized the data of Faber \etal\ 1989 and
Dressler 1987), and J{\o}rgensen \etal\ (1995b, who increased the JFK
sample and obtained average velocity dispersions from a fixed and a 
variable, within $\re/4$, aperture of the slit).

The accuracy of the kinematical data is more difficult to evaluate.
Much of the offset in the distance determinations are likely caused
by systematic differences in the $\sigma$ measurements. Such differences
are difficult to evaluate due to the heterogenous nature of the samples and
of the methods adopted to gauge the central velocity dispersion.

According to Jacoby \etal\ (1992) the errors on $\sigma$ likely range
between 6 and 14\%. It is clear that such an estimate does not account for
possible local deviations from the overall dynamical behaviour due for
instance to singularities, which may cause the appearance of central
spikes in the velocity dispersion profile (which behave as systematic errors on
$\sigma$).

Of the 52 early--type galaxies in the Virgo sample (29 S0 and 23 E), 
13 S0s and 4 Es
have no $\sigma_0$, 3 S0s have $\sigma_0 < 100$ \kms, and 4 S0s and 2 Es
have anomalous high/low redshifts with respect to the mean of the cluster
($cz_0=1179\pm17$, Sandage \& Tammann 1995). One galaxy (NGC\,4370) has been
eliminated because its central surface brightness is fainter than
$\mu_B=20.75$ (implying $\dn=0$). In Fornax we have 12 S0s and 16 Es: 7
S0s and 7 Es have no measurements of $\sigma_0$, and 2 S0s have 
$\sigma_0 <100$ \kms.
There are no peculiar redshifts.

\subsection{The \dns\ relation for the Virgo and Fornax clusters}\label{sec3b}
Fig.~\ref{fig1} plots the \dns\ relation for the \CCD\ sample. The upper
panel shows only the Virgo galaxies, with different symbols according to
their morphology: open circles for Es and triangles for S0s. Here
the central velocity dispersions of the Virgo galaxies are the plain
averages of the data found in the literature.

It is apparent that E and S0 galaxies follow the same relation. There are,
however, 8 Virgo galaxies that largely deviate from the mean relation
(indicated by the solid line; see below).

The discrepant galaxies have peculiar redshifts and/or high inclination to
the line of sight. Three of them have $\sigma_0 < 100$ \kms\ and,
following the recipe of the 7-Sam have been discarded.
Of the 9
S0s that follow the mean relation, 3 have apparent flattening higher than
S0(5) (according to the BST classification) and one of them has a
peculiar redshift. Two galaxies fitting the mean relation are rounder than
S0(5) but have discrepant redshifts. Therefore it seems that inclination
and redshift do not determine univocally the galaxies that follow
the \dns\ relation and are not useful selection criteria for establishing a
``representative'' sample.

Table~3a presents the morphological type, the axis ratio, the recession
velocity and the BST membership code for the Virgo galaxies which deviate
significantly from the mean \dns\ relation, and Table~3b the same data for
those S0s that instead fit the relation. In the following we will reject
the 8 discrepant galaxies when computing the \dns\ and FP relations (the
same galaxies, with the possible exception of NGC\,4200, also show a large
scatter with respect to the FP); they are likely foreground and background
objects.

The solid line in Fig.~\ref{fig1} has been computed through a linear
least-square fit of the best Virgo sample, ignoring errors
on both coordinates. We get: 
\begin{equation}\label{eq1} 
\log(D_n) = (1.26\pm0.06) \log(\sigma_0) - (1.11\pm0.14) 
\end{equation} 
with a \rms\ of 0.06 in $\log (D_n)$, corresponding to a 15\% uncertainty in 
the average distance.

With a more robust fit which takes into account the errors on both
coordinates (Fasano \& Vio 1988), it is:
\begin{equation}\label{eq2} 
\log(D_n) = (1.31\pm0.07) \log(\sigma_0) - (1.24\pm0.16)
\end{equation} 
and the same \rms\ fit as above\footnote{Note that the errors on both 
coordinates ($\Delta\log(D_n)\sim0.02$ and $\Delta\log(\sigma)\sim0.04$) are
approximately equal, and the use of a least square fit that  
weights both variables does not produce a significantly different result.}.

Taking into account the above quoted errors, the intrinsic scatter of the 
\dns\ relation is 0.05, corresponding to a distance uncertainty of $\sim12\%$ 
per galaxy.

In the lower panel of Fig.~\ref{fig1} the Virgo and Fornax galaxies are
indicated with different symbols. The solid line is our eq.~\ref{eq1}.

\begin{table}
\begin{tabular}{llccc}
\multicolumn{5}{c}{Table 3a: Virgo galaxies that do not fit the \dns\
relation} \\
\hline\hline \\[-7pt]
Ident. & Morph. & $\displaystyle\left(\frac{b}{a}\right)_{25}$ & $cz$ &
Membership \\  & Type & & [$\kms$] & \\
\hline \\[-7pt]
NGC\,4168 & E2    & 0.79 & 2316 & -- \\
NGC\,4200 & S0(4) & 0.61 & 2376 & -- \\
NGC\,4261 & E2    & 0.80 & 2200 & -- \\
NGC\,4270 & S0(6) & 0.46 & 2347 & -- \\
NGC\,4281 & S0(6) & 0.56 & 2711 & -- \\
NGC\,4342 & S0(7) & 0.54 & \ph{2}714 & -- \\
NGC\,4417 & S0(7) & 0.62 & \ph{2}832 & M \\
NGC\,4550 & S0(7) & 0.31 & \ph{2}378 & M \\
\hline \\ 
\end{tabular}
\end{table}

\begin{table}
\begin{tabular}{llccc}
\multicolumn{5}{c}{Table 3b: Virgo S0s fitting the \dns\ relation } \\
\hline\hline \\[-7pt]
Ident. & Morph. & $\displaystyle\left(\frac{b}{a}\right)_{25}$ & $cz$ &
Membership \\  & Type & & [$\kms$] & \\
\hline \\[-7pt]
NGC\,4339 & S0(0) & 0.86 & 1287 & -- \\
NGC\,4377 & S0(3) & 0.85 & 1371 & M \\
NGC\,4459 & S0(2) & 0.84 & 1210 & M \\
NGC\,4526 & S0(6) & 0.37 & \ph{2}533 & M \\
NGC\,4552 & S0(0) & 0.85 & \ph{2}321 & M \\
NGC\,4570 & S0(7) & 0.31 & 1730 & M \\
NGC\,4578 & S0(4) & 0.70 & 2284 & M \\
NGC\,4638 & S0(7) & 0.70 & 1147 & M \\
NGC\,4649 & S0(2) & 0.81 & 1144 & M \\
\hline 
\end{tabular}

\footnotesize{
Notes to Tables 3a,b: Morphological types, recession velocities and
membership are from Binggeli \etal\ (1985); axis ratios at $\mu_B=25$ are
from \CCD.}
\end{table}

If we assume that the \dns\ relation for the Fornax cluster has the same slope 
as that for Virgo (a reasonable approximation if the environment does 
not influence the \dns\ relation), the fit providing the minimum scatter for 
the Fornax galaxies gives $\Delta\log(D_n)=0.09$, 
and a ratio $D_n\mbox{(Virgo)}/D_n\mbox{(Fornax)} = 1.23$, corresponding to 
$\Dmufv = 0.45$ mag.

This result disagrees with Davies \etal\ (1993), who found
$D_n({\rm Virgo})/D_n({\rm Fornax}) = 1.07$, and a $\Dmufv = 0.15$ mag. 
The origin of such a discrepancy (15\% in distance) is not in the values 
of \dn, since the two samples compare quite well, but rather
in the values of $\sigma$. For 7 of the 8 Fornax 
galaxies we have in common, the values of $\sigma_0$ are larger in our sample. 
The discrepancy between the spectroscopic data of the two samples
increases for the objects with $\sigma_0 > 150$ \kms\ (D'Onofrio \etal\
1995). 

The non-homogeneity of the velocity dispersion data is also clear from
Fig.~\ref{fig1}. Note that the scatter for the Virgo objects is lower than
for Fornax: the \rms\ of the least square fit of the two clusters 
is 0.06 for Virgo and 0.13 for Fornax\footnote{Excluding NGC\,1316, the 
galaxy that has the largest discrepant value of $\sigma_0$, the \rms\ 
lowers to 0.10.}.

\begin{figure*} 
\psfig{figure=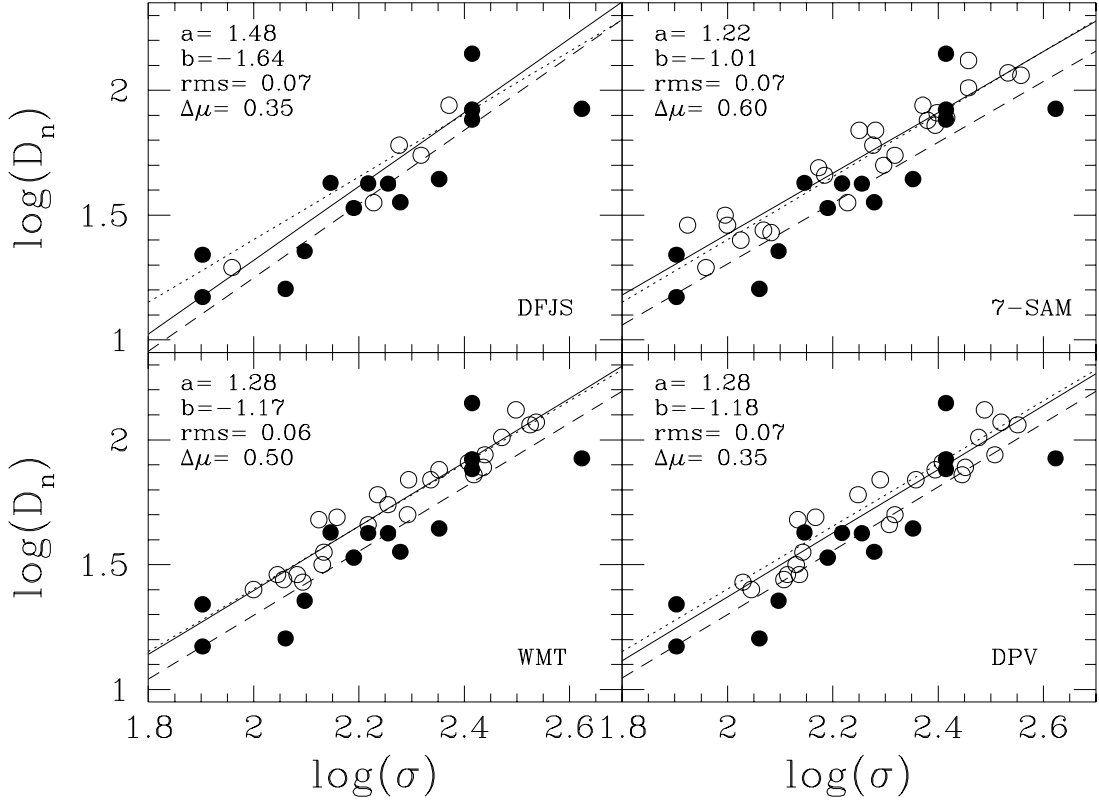,width=17cm,angle=270}
\caption[]{The \dns\ relation using individual sources of $\sigma_0$ for
the Virgo galaxies. 
Open and filled symbols indicate Virgo and Fornax objects respectively. 
The solid lines are the fit the Virgo data, for which we report intercept 
$a$, slope $b$ and \rms. The dotted line is our eq.~\ref{eq1}. 
The dashed lines are the fit for Fornax, assuming the same slope of Virgo, 
from which \Dmu\ is computed.} 
\label{tutti}
\end{figure*}

This is at variance with the expectation based on the morphology of the
two clusters: diffuse (Virgo) and compact (Fornax), according to BST
and Ferguson (1989).
If one assumes that the two clusters have spherical symmetry, an intrinsic
$\sim7\%$ and $\sim11\%$ scatter is expected for Fornax and Virgo respectively,
according to their angular sizes.

One might speculate that the small scatter of the Virgo data is due to the
averaging of several measurements of $\sigma$, obtained by different authors
using different methods of data analysis on spectra taken with different slit
apertures. 
What does it happen if we do not average the central velocity dispersion
measurements\,? 
This is shown in Fig.~\ref{tutti} where, for each source of $\sigma_0$ data, 
we report the slope, intercept, and \rms\ scatter of the best fit 
for the corresponding Virgo galaxies and the resulting \Dmufv\ computed 
adopting the slope derived for Virgo.

This experiment tells us two things. The first is that the small scatter
of the Virgo data is independent of the databases (each catalog of velocity 
dispersions is made of averaged and homogenized data). 
If we use, however, the 12 galaxies of the sub-sample with the best observed 
rotation curves and velocity dispersion profiles, the \rms\ scatter from the 
best fit increases to 0.08 (20\%). 
The second is that the relative distance moduli between the two
clusters span a large interval, varying with the properties of the
adopted Virgo sample. The null value of \Dmufv\ found in our
comparison with Pierce (1989) demonstrates this effect. 

Since most of the Virgo data come from the 7-Sam sample, we decided to
concentrate on the errors introduced by the aperture sizes on the relative
distance between Virgo and Fornax, adopting the central velocity dispersions
of their Lick and LCO-HI measurements ($4'' \times 1''.5$ and $4'' \times 4''$
slit aperture respectively). The two data-sets for precisely the same
galaxies, give \Dmufv\ $=$ 0.35 and 0.55 respectively. This means that an
accurate distance between the clusters could be obtained only after an
homogenous comparison of the $\sigma$ measurements, corrected for the aperture
size effects. A more quantitative idea of such corrections has been provided
by J{\o}rgensen \etal\ (1995b). 

Finally, we calculate the distance between Fornax and Virgo 
through the $L - \sigma$
relation using the \CCD\ $B$-magnitudes for the galaxies of the two clusters
and the velocity dispersions from the catalog of McElroy (1995).
We obtained $\Dmufv = 0.43$ mag, in good agreement with the \dns\ relation.

\subsection{The \dns\ relation for the Coma cluster}\label{sec3c}
We used the data given by LGCT, JFK and J{\o}rgensen \etal\ (1995a,b) to derive
the distance to the Coma cluster. Unfortunately, the \dn\ diameters measured
by LGCT are in the $V$ band (instead of $B$), at the mean surface brightness
level of 19.8 \mga. However, since the average $B-V$ colour of elliptical
galaxies is approximately constant and equal to 0.95 mag, we can equally
compare the two data sets. The same happens for the \dn\ values of 
J{\o}rgensen 
\etal\ (1995a) which are in the $r$ band. In that case, using the photometric
parameters in the $B$ and $r$ bands listed in JFK, we derived the relations:
$\log(r_n^B) = 0.968\,\log(r_n^r) + 0.017$ and
$\log(r_e^B) = 1.006\,\log(r_e^r) + 0.026$.
The new parameters of J{\o}rgensen, together with those given by LGCT, 
have been corrected by the authors for cosmological and seeing effects.

\begin{figure} 
\psfig{figure=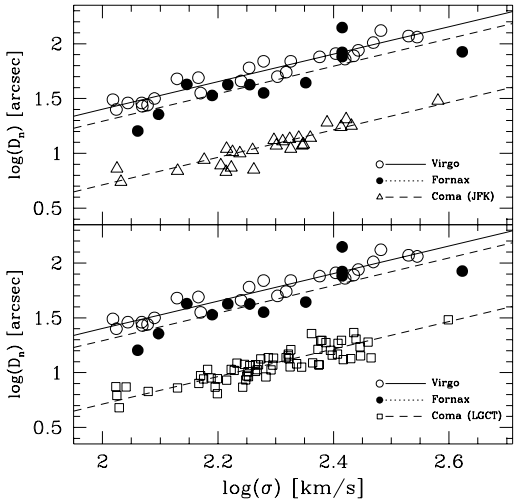,width=8cm,angle=0}
\caption[]{{\it Panel a)} The $\log(D_n) $-$\log(\sigma)$ relation for the
Virgo, Fornax, and Coma samples using the data from
LGCT for the latter cluster. {\it Panel b)} same as panel {\it a}
with the data of JFK. The solid, dotted, and dashed lines give
the fits to the Virgo, Fornax, and Coma data-points respectively.} 
\label{fig2}
\end{figure}

For the LGCT and JFK samples the line of minimum scatter with respect to
eq.~\ref{eq1} (Fig.\ref{fig2}) gives 
$\Delta\log(D_n)=0.68$ and $\Delta\log(D_n)=0.69$ respectively,
corresponding to a \Dmucv\ of 3.4 and 3.45 mag.
Considering a 5\% aperture correction to the observed $\sigma_0$ passing
from the
Virgo and Coma galaxies (as in Davies \etal\ 1987), 
\Dmucv\ turns out to be $\sim3.55$ mag.
We note that the use of the LGCT seeing corrected parameters
does not influence very much the final result.

We also calculated the relative distance between Coma and Virgo using the
photometric parameters and the Coma velocity dispersions quoted by
J{\o}rgensen \etal\ (1995b), who give the average $\sigma$ within a fixed
aperture of $3''.4$ and within a variable aperture $r_e/4$, which is connected
to the physical size of the galaxies. It was impossible for us to get the same
variables for the Virgo galaxies, since this requires the velocity dispersion
profiles. We found $\Dmucv=3.45$ and 3.55 mag respectively in the two cases,
confirming the 12-15\% distance uncertainty due to aperture effects. Since the
$r_e/4$ aperture is generally lower than the $3''.4$ slit for the Coma
galaxies, we conclude that, by enlarging the aperture, one obtains a lower
average value of $\sigma$ and consequently a lower differential distance
modulus. 

It is also interesting to compare the \dns\ results with the 
$V_{26}-\sigma$ relation of Dressler (1984, hereafter AD), who found $\Dmucv =
3.75$ mag. His result is particularly important here because Dressler
considered two different slit apertures for the galaxies in Virgo and Coma in
order to reduce as much as possible the aperture effect. The $\sim0.25$ mag
difference of the two methods can be explained either by a systematic
difference in \dn\ (15\%) or in $\sigma_0$ (10\%). 

We first compared the magnitudes of the Virgo galaxies in common
with Dressler (he used the data of Sandage \& Visvanathan 1978).
>From the growth curves of the \CCD\ sample, we derived the $B_{26}$ 
magnitudes of our galaxies, and compared them
with Dressler's data by transforming 
his $V_{26}$ magnitudes in the $B-$band. We adopted the color equation
of Sandage \& Visvanathan (1978): $B-V = 1.25 (b-V) + 0.22$.
We obtained $B_{our} = 1.03\,B_{Dre}-0.80$. 
Being only a zero point offset in the $B$ magnitudes, the resulting
$B_{26} - \sigma$ relation for the Virgo-Coma galaxies
produces a \Dmucv\ of 3.80.
Unfortunately, Dressler does not report the \dn\ of his galaxies,
and so a definitive comparison cannot be made. However, when we compared
our data with Dressler's \etal\ (1987), we found a good agreement. This
is an indication that the observed discrepancy is not
due to a systematic variation of the photometric parameters.

The discrepancy between the \dns\ and the $L-\sigma$ relations can also
be due to a systematic $\sim10\%$ offset in the velocity
dispersions. Part of this offset ($\Delta\log(\sigma)=0.028$) has been
found by LGCT in their comparison with the data of Dressler
(unfortunately however, such a correction to the $\sigma$ values of Dressler
brings the result of the $B-\sigma$ relation to $\Dmucv = 4$).
By comparing the central velocity dispersions of the Coma galaxies from
LGCT, Dressler (1984), and McElroy (1995, hereafter McE),  
we found $\Delta\log\sigma_0(McE-LGCT)\sim-0.017$ and
$\Delta\log\sigma_0(McE-AD)\sim0.014$. 
For the Virgo galaxies we made the comparisons between Dressler (1984),
McElroy (1995) and Davoust \etal\ (1985, herefater DPV) obtaining
$\Delta\log\sigma_0(McE-AD)\sim0.015$ and
$\Delta\log\sigma_0(DPV-AD)\sim0.04$ (note that the DPV data 
have been homogeneized by the authors taking into account aperture
effects, method of reduction, etc.).
This means that the
aperture effects can account for the observed difference.
The large differential distance modulus is likely the result of the 
low central velocity dispersion
that Dressler got for the Virgo galaxies.

Looking at the data of the 7-Sam and of DPV, 
we discovered that in the $70\div75\%$ of the cases the use of a small 
aperture (slit width over face on corrected diameter $D(0)$ less than 0.02)
produces an higher central velocity dispersion ($\sigma_l/\sigma_s=0.85$), 
where the subscripts here stay for "large aperture" and "small aperture". 
This effect can be easily
responsible of the scatter observed in \Dmucv, since it gives $\sim$15\%
difference in the central velocity dispersions.

A further possible explanation of the difference between the $L - \sigma$
and \dns\ relation resides in the surface brightness bias which
affects the $L - \sigma$ relation. To test this effect we corrected the
total $B$ magnitudes of the galaxies adopting the relation 
$\Delta B = -0.3 \muem + 6.49$, which
represents the trend of the residuals as a function of the effective surface
brightness in our data. The result is that we obtain a change in the slope and
in the scatter of the $B - \sigma$ relation, but the differential modulus is
not affected. 

The distance to Coma derived here is in conclusion slightly smaller than 
the results of other DIs. 
The responsibility of this difference is likely in the adopted values of
the central velocity dispersions.
Within the uncertainty, however, the adopted value is in good agreement 
with the B--band TF relation (Fukugita \etal\ 1993), 
with the \dns\ relation of Dressler \etal\ (1987),
with the SN--Ia (Capaccioli \etal\ 1990), and with Bower \etal\ (1992), who
used a combination of three methods: CM diagram, colour-diameter
relation, and FJ relation.

Another possible source of error, discussed by LGCT
and Lucey \etal\ (1991a), is the surface brightness bias which may affect
the \dns\ relation of the Coma galaxies. 
Such a bias is evident in the LGCT
data looking at the residuals of the \dns\ relation for
the galaxies in the central region of the Coma cluster with respect to those
in the outer parts.
We checked the influence of this bias on
\Dmucv\ by selecting objects in the core of the cluster and outside,
but we did not find any significant difference between the two samples.

\subsection{The aperture effects on the central velocity 
dispersion}\label{sec3d}
Since aperture effects play a major role in determining the measured central
velocity dispersion, we perform another test to better investigate this
problem. 

\begin{figure} 
\psfig{figure=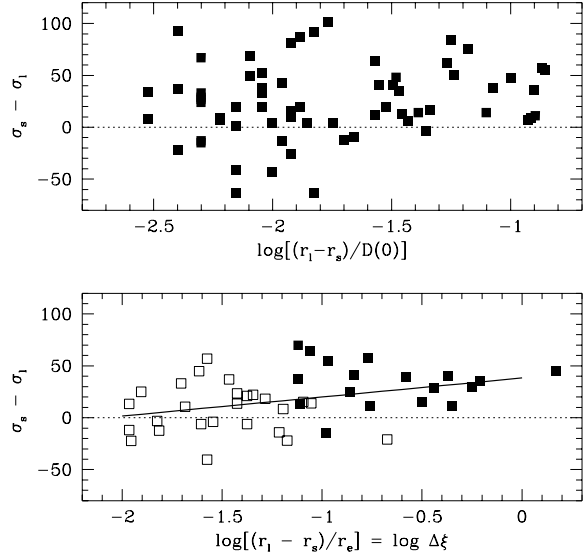,width=8cm,angle=0}
\caption[]{{\it Bottom panel)} The open square represent the difference between
the velocity dispersions of the Virgo galaxies 
obtained through the slits of 2 and 4 arcsec 
respectively. The filled square that from 2 and 16 arcsec.
The solid line is a fit to the galaxies of our Fornax sample (see text).
{\it Upper panel)} The differences of $\sigma$ from small and large apertures
from the catalog of Davoust \etal\ (1985).} 
\label{aperture}
\end{figure}

In Fig.\ref{aperture} (bottom panel) we plot the difference between the 
velocity dispersion values for the same galaxy measured through two
apertures of different radius $r_s < r_l$ (small and large slit apertures)
versus the aperture size increment in units of the galaxy 
effective radius. 
The data are for the Virgo galaxies: the velocity
dispersions have been measured from slit widths of $2''$, $4''$ and $16''$ 
and are taken from the 7-Sam and Dressler (1984) samples.
The solid line represents a fit obtained for the
galaxies of our Fornax sample. For each galaxy we calculate the difference
between the central velocity dispersion and the luminosity
weighted velocity dispersions within increasing fractions of the effective
radius. The average values for 13 galaxies can be fitted by the
relation:
$\sigma_0 - \sigma_{ap} = 18.3 \times \log(r_{ap}/\re) + 38.3$ with a
correlation coefficient of 0.97 and a rms of 2.57. 

The upper panel reports the data presented in the catalog of Davoust \etal\
(1985). Here the authors listed the width of the slit normalized to the 
corrected face-on diameter of the galaxies $D(0)$.

The figure shows that when the difference among the apertures is larger
than $\sim0.1$ \re, 
the value of $\sigma$ obtained through the small aperture is systematically
larger than that given by the large slit. It should be emphasized that for the
smaller galaxies this effect is much more evident (the slit covers a larger
fraction of the effective radius).

The scatter around zero at smaller $\Delta\xi=(r_l-r_s)/r_e$ is possibly
explained by the 10-15\% error of the individual measurements, but it may have
also another explanation. 
According to Ciotti \etal\ (1996) and Ciotti \& Lanzoni
(1996), the projected central velocity dispersion depends
on the fraction of effective radius of the galaxy covered by the slit width
and on the surface brightness distribution of the galaxy as whole.
The ratio of the central to the virial $\sigma$ as a function of $r/r_e$ has
approximately a gaussian shape, with a peak centered between $0.1\div0.3\re$.
The shape and position of the peak depend on the exponent $m$ of the
$r^{1/m}$ law, which best fits the light distribution of early-type galaxies
(Caon \etal\ 1993). When $m$ is greater than 4, the curve becomes progressively
monotonic decreasing. 
This particular behaviour can be responsible of the observed scatter around 
zero in Fig.\ref{aperture}. In fact, according to the fraction of \re\ covered
by the slit one can get different $\sigma$.
Only for the galaxies with $m>4$ the difference between the central velocity
dispersion obtained from the small ($\sigma_s$)  and large ($\sigma_l$) 
apertures is always greater than zero.
Interestingly enough we verified that for the 
Virgo galaxies with $m>4$ we always have $\sigma_s - \sigma_l > 0$.
Anyway, taking into account the uncertainties on the measured $\sigma$,
the prediction of this model cannot be presently confirmed.

By applying the aperture correction to the data of Dressler (1984) for the
Virgo and Coma clusters the resulting \dns\ relation is $\log(D_n) = 1.53
\log(\sigma) - 1.79$ and the differential distance modulus \Dmucv $=$
3.685.
The same correction applied to
the data of the 7-Sam (for Virgo) and to our Fornax sample gives 
$\log(D_n) = 1.53 \log(\sigma) - 1.81$ and \Dmufv $=$ 0.35. Notice the
change of the slope from 1.26 to 1.53, and the better agreement of the
differential distance moduli with the average literature values.

\subsection{The error analysis}\label{sec3e}
As already stated, the limits of the above results in the determination of
the differential distance moduli are in: 1) the {\it a
priori\/} assumption that the slope of the \dns\ relation is the same for
the three clusters, 2) the choice of the ``representative'' sample of
objects for the whole cluster, 3) the accuracy of the data (in particular
$\sigma_0$), and 4) the definition of the variables which enter the relation:  
\eg\ $\sigma_0$ or the average $\sigma$ within a prefixed radius.

\begin{figure} 
\psfig{figure=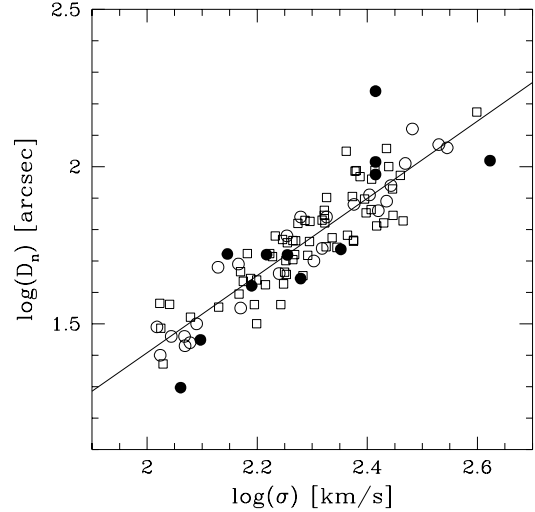,width=8cm,angle=0}
\caption[]{The fit to the full data-set of Virgo, Fornax, and Coma
galaxies (\CCD, LGCT, and JFK samples, for a total of 128 objects), after
shifting Fornax and Coma to the distance to Virgo. Symbols are the same as
in the previous figures.}
\label{insieme}
\end{figure}

Before turning our attention to the errors associated to the use of the
\dns\ relation, we have checked the assumed invariance of the slope
$d\log(D_n)/d\log\sigma_0$. To this end we have shifted the Fornax and
Coma galaxies to the distance to the Virgo cluster using the above
calculated $\Delta\log(D_n)$, and fitted the whole distribution again.
The result:
\begin{equation}\label{eq3}
\log(D_n) = (1.24\pm0.06)\,\log\sigma_0 - (1.08\pm0.15),
\end{equation}
with \rms\ $=$ 0.08,
shown in Fig.~\ref{insieme}, is in excellent agreement with eq.~\ref{eq1}.

Let us now analize the fitting procedure. Isobe \etal\ (1990) and
Feigelson \& Babu (1992) developed five methods for applying linear
regression fits to a bivariate data distribution with unknown or
negligible measurement errors. The advantage of their method is that the
uncertainties in the slope and intercept of the fitted distribution, and
in the process of comparative calibration, are evaluated through resampling
procedures: jackknife and bootstrap error analysis. The bootstrap error
analysis is based ``on the distribution of slopes and intercepts of a
large number of data sets constructed by random sampling of the observed
data set with replacement''. The jackknife error analysis is performed in
a similar way but with only $N$ synthetic data sets, each containing
$(N-1)$ points from the original data set, leaving out one observation in
sequence.

Following their approach we used the standard linear regression analysis
(program SLOPES, kindly provided by the authors),
with the distance dependent variable ($\log(D_n)$) as the Y variable. This
choice has the effect of minimizing the residuals in \dn. The slopes for
the Virgo, Fornax, and Coma samples range from 1.16 to 1.37, with
different errors for the different samples. For the Fornax cluster the
errors in the intercept (0.69) and in the slope (0.31) are much larger than
for the other two clusters, due to the large scatter of the data and to the
small number of galaxies.

Which is the error that affects the intercept offset\,? Assuming that the
distance to the calibrating cluster is known with high accuracy (\ie\ that
intercept and slope of the fit are error free), the error in the relative
distance comes approximately from the uncertainties in $\log(D_n)$
($E_D=\Delta\log(D_n)$) and in $\log\sigma$ ($E_\sigma=\Delta\log(\sigma)$) 
for the new sample of galaxies. Since we are looking for the fit that
minimizes the residuals of the new sample, and we assume that the slope is
the same as that of the calibrating sample, we can set the maximum shift in
the intercept equal to $\sim\sqrt{E_D^2 + E_\sigma^2}$. Being the 
errors in both coordinates equal to 0.02 (minimum error), the minimum shift
of the intercept is $\sim0.03$, corresponding to an error in \Dmu\ of
0.14 mag or 6\% in distance.

A more correct estimate of such errors comes from the program CALIB
developed by Isobe \etal\ (1990) and Feigelson \& Babu (1992), which
predicts the value of the intercept (and its uncertainty) in the fit of a
new sample of galaxies, from the linear regression of a calibrating
sample, without the assumption that the two slopes are equal.

In summary this procedure gives $\Dldnfv=0.093\pm0.015$ and
$\Dldnfv=0.095\pm0.06$, assuming Virgo and Fornax respectively as
calibrators, and $\Dldncv=0.693\pm0.013$ and $\Dldncv=0.690\pm0.016$ for
Coma and Virgo, A 0.015 uncertainty in $\log(D_n)$ corresponds to
$\sim3\%$ error in the relative distance to the clusters. This is the
minimum uncertainty attainable with such data for the relative distance
between the two clusters.

\section{The Fundamental Plane relation}\label{sec4}
\begin{figure} 
\psfig{figure=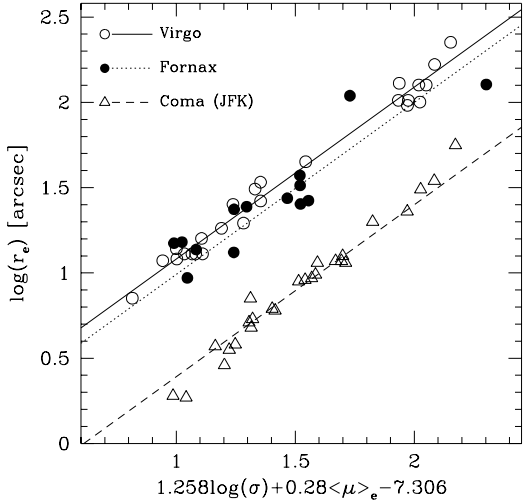,width=8cm,angle=0}
\caption[]{The FP for the Virgo (open circles), Fornax (filled circles),
and Coma (open triangles) galaxies. The solid line is our eq.~\ref{eq4},
the dotted and dashed lines represent the fit of minimum scatter for
Fornax and Coma respectively, assuming the same slope of Virgo.} 
\label{fp}
\end{figure}

For the Virgo galaxies of the \CCD\ sample, the FP relation shown in
Fig.~\ref{fp} writes: 
\begin{eqnarray}
\log(\re) &\!\!=\!\!& (1.26\pm0.09)\log(\sigma) +
\nonumber\\
&& (0.28\pm0.01)\muem\ - (7.31\pm 0.28).
\label{eq4}
\end{eqnarray}
The values of the coefficients are derived here through a multivariate
statistical analysis. The \rms\ error of the fit is 0.06, which translates
into a 15\% uncertainty in the distance to a single object, the same
obtained with the \dns\ relation. For comparison JFK found for the Coma
cluster: $\log(\re)=(1.20\pm0.09) \log(\sigma)+(0.35\pm0.01) \muem\ -
(9.31\pm0.36)$, with a \rms\ of 0.046.

Assuming that the slope provided by the Virgo cluster is also valid for
Fornax and Coma, and using the JFK data we found respectively
$\Dlrefv=0.10$ and $\Dlrecv=0.69$. 
Using the seeing corrected parameters of J{\o}rgensen \etal\ (1995a) we found
a $\Dmucv = 3.5$ mag in good agreement with the above uncorrected quantities.
 
This means that the FP method gives exactly the same relative distance 
to Virgo obtained through the \dns\ relation. 
The same result is obtained for Coma if we adopt the FP relation derived by
JFK.

Now, by setting $X = 1.26\,\log(\sigma)+0.28\,\muem\ - 7.31$ and
$Y=\log(\re)$, and running again the program SLOPES and CALIB, we obtain:
$\Dlrefv=0.06\pm0.01$ and $\Dlrefv=0.07\pm0.05$, using Virgo and Fornax
respectively as the calibrating samples, $\Dlrecv=0.69\pm0.01$ and
$\Dlrecv=0.70\pm0.02$, using instead Virgo and Coma. Again the
uncertainty in $\log(\re)$ corresponds to a $\sim3\%$ error in the relative
distance to the clusters.
The results of these programs are summarized in Table~4.

If we shift Fornax and Coma to the distance of Virgo we obtain for the 
whole sample of 65 galaxies: $\log(\re) = 1.02 \cdot X + 0.07$ with \rms\ $=$ 
0.09, comparable to the intrinsic scatter of the relation, and
corresponding to an uncertainty of $\sim$23\% in distance. 
This is a value somewhat larger than previously reported in the literature.

The distance moduli to Fornax and Coma relative to Virgo, based on the FP
relation, now span the interval $0.3 \leq \Dmufv \leq 0.45$ mag, and $3.45
\leq \Dmucv \leq 3.50$ mag, depending on the fitting method that is
used.

\section{Testing the \dns\ and FP relations}\label{sec5}
The second part of this paper is dedicated to a critical analysis of the
\dns\ and FP relations, taking advantage of our well defined (homogeneous
and volume-limited) sample of galaxies in Virgo and Fornax.

\subsection{The surface brightness bias in the \dns\ relation for the
Virgo and Fornax clusters}\label{sec5a}
The existence of a surface brightness bias in the \dns\ relation has been
claimed by many authors (\eg\ Lynden-Bell \etal\ 1988, LGCT, JFK, van
Albada \etal\ 1993). We have looked for such a bias using the \dns\
relation modified by adding a term in surface brightness, as proposed by
van Albada \etal\ (1993):
\begin{equation}
\label{eq5}
\log(D_n)=a\,\log(\sigma) + b\,\Delta\mu + c\,\Delta\mu^2,
\end{equation}
where $\Delta\mu = \muem - 20.75$.
For our Virgo sample we obtain $a=1.356$, $b=-0.008$, and $c=0.002$, with
a \rms\ scatter of 0.06. The multivariate statistical analysis performed
on the three variables $\log(D_n)$, $\log(\sigma)$, and $\Delta\mu$, shows
that 95\% of the total variance is enclosed in the correlation between the
first two variables. For comparison the coefficients found by van Albada
\etal\ (1993) with the data of Faber \etal\ (1989) are $a=1.35$,
$b=0.047$, and $c=-0.019$. With our data the contribution of the terms in
surface brightness is very low.

\begin{figure} 
\psfig{figure=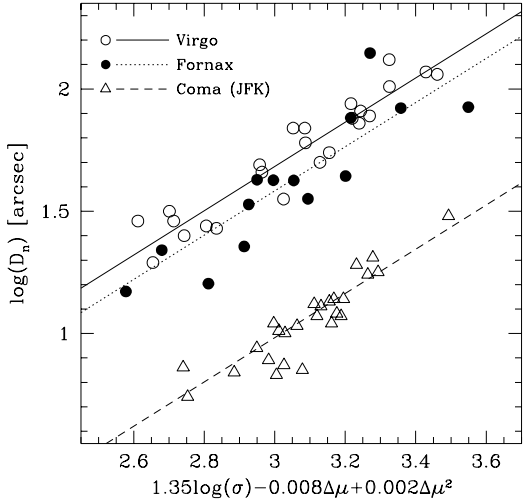,width=8cm,angle=0}
\caption[]{The \dnsmu\ relation for the Virgo, Fornax, and Coma samples of
galaxies. Symbols are the same as in the previous figures. The solid line
is our eq.~\ref{eq5}, the dotted and dashed lines represent the fit of
minimum scatter for Fornax and Coma, assuming the same slope of Virgo.}
\label{figdmu}
\end{figure}

We show in Fig.\ref{figdmu} the distribution of the galaxies of the \CCD\
and JFK samples in the \dnsmu\ diagram. Again the offsets
among the clusters distributions are of the same order as in the classical
\dns\ and FP relations.

\begin{figure*} 
\psfig{figure=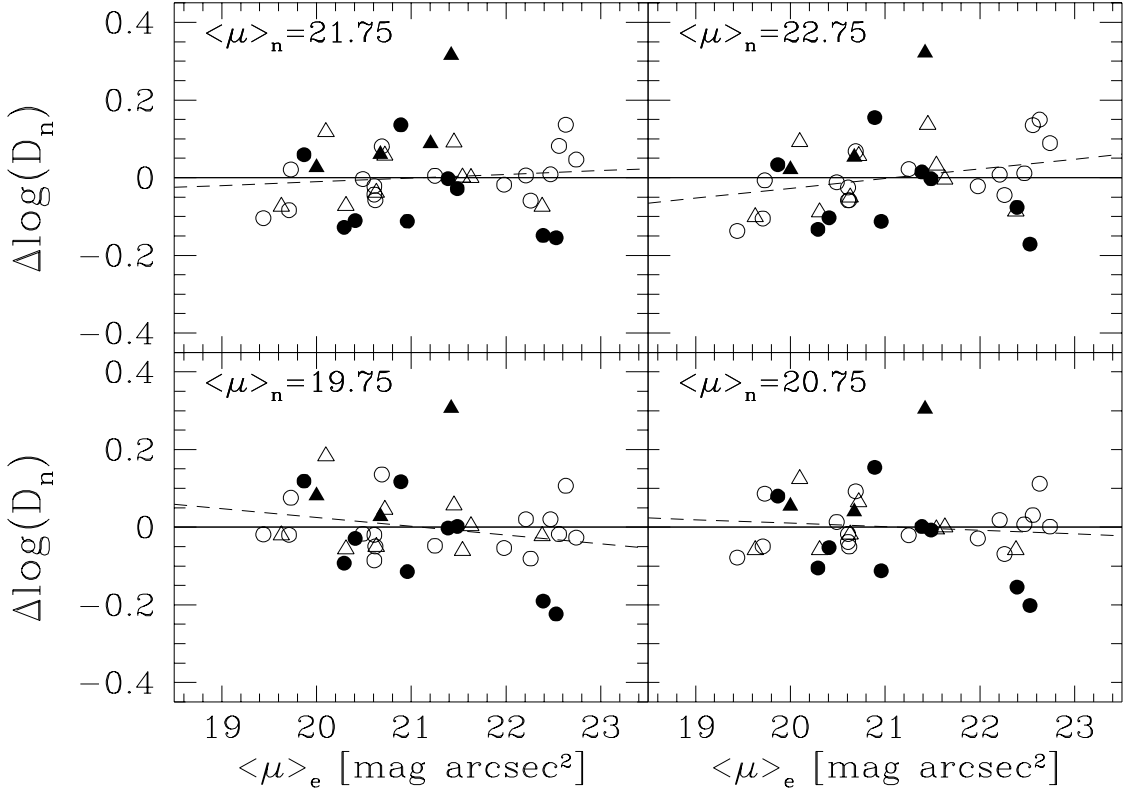,width=17cm,angle=270}
\caption[]{Residuals of the \dns\ relation versus the average surface
brightness \muem\ for the galaxies of the \CCD\ sample. 
Different choices of \muen\ are used in each box (see text).
Open and filled symbols represent Virgo and Fornax objects respectively.
Circles and triangles are E and S0 galaxies respectively. The dashed 
lines are best fits to the plotted distributions.} 
\label{res}
\end{figure*}

The trend of the residuals $\Delta\log(D_n)$
in the \dns\ relation with respect to \muem\ is
shown in Fig.\ref{res} for the galaxies of the \CCD\ sample. Here we have
explored different definitions of \dn\ by varying the average surface 
brightness \muen\ from 19.75 to 22.75 \mga. It is apparent
that there is no clear trend of the residuals with \muem, in agreement
with LGCT, but not with JFK. The \rms\ scatter of the residuals is similar
in all cases and equal to 0.09.

Following Dressler \etal\ (1987), who showed that $\dnae\propto\Iem^{0.8}$
for \rq\ galaxies, D'Onofrio \etal\ (1994; their eq.~8 and Fig.~1) derived
the relation between \dnae\ and \muem\ for objects with \rn\ light
profiles. Here $m$ is the exponent that provides the best fit to the
major axis light profiles for the early--type galaxies of the \CCD\
sample, and ranges from 1 to $>10$, increasing systematically with 
the galaxy luminosity.

Under simple assumptions, the inverse relation $\muem - \muen$ versus
\dnae\ can be written:
\begin{equation}
\label{eq6}
\log(\dnae) = \sum_{k=0}^{\infty} a_k(m) [\muem - \muen]^k
\end{equation}
where \muen\ is the average surface brightness within a prefixed isophote
(in general \muen $= 20.75$), and the coefficients $a_k(m)$ are functions
of $m$.

Writing the observed \dns\ relation as:
\begin{equation}\label{eq7}
\log(\dn) = a'\,\log\sigma + b'
\end{equation}
and the FP equation as:
\begin{equation}
\label{eq8}
\log(R_{\rm e}) = a\,\log\sigma + b\langle\mu\rangle_{\rm e} + c,
\end{equation}
the difference between eqs.\ref{eq6} and \ref{eq7}
can be expressed by the formula:
\begin{eqnarray}
\Delta\log(\dn) &\!\!=\!\!& [a-a']\log\sigma+[b+a_1(m)]\muem+
\nonumber\\
&&\sum_{k=2}^{\infty} a_k(m) [\muem - \muen]^k + \mbox{const},
\label{eq9}
\end{eqnarray}
where const=$a_0(m)+c-b'-20.75\,a_1(m)$.
The values of the first three coefficients $a_k$ for $m=4$ are approximately:
$a_0=1.2\times10^{-3}$, $a_1=-2.7\times10^{-1}$, $a_3=-1.5\times10^{-2}$,
vanishing for large values of $k$.

Looking at the values of the slope and of the intercept of the \dns\ and FP 
relations, one sees that all the coefficients in eq.~\ref{eq9} are very small 
(\eg\ $[a-a']\sim 0.05$, assuming respectively the slopes in eq.~\ref{eq3} and
eq.~\ref{eq4}, $[b+a_1(4)]= 0.007$, and so on) for each value of $m$. For
this reason the dependence on \muem\ of the residuals of the \dns\ is
almost completely cancelled out, and this is confirmed also in our
Fig.~\ref{res}.

\subsection{The behaviour of the residuals in the \dns\ and the Fundamental 
Plane relations}\label{sec5b}
\begin{figure*} 
\psfig{figure=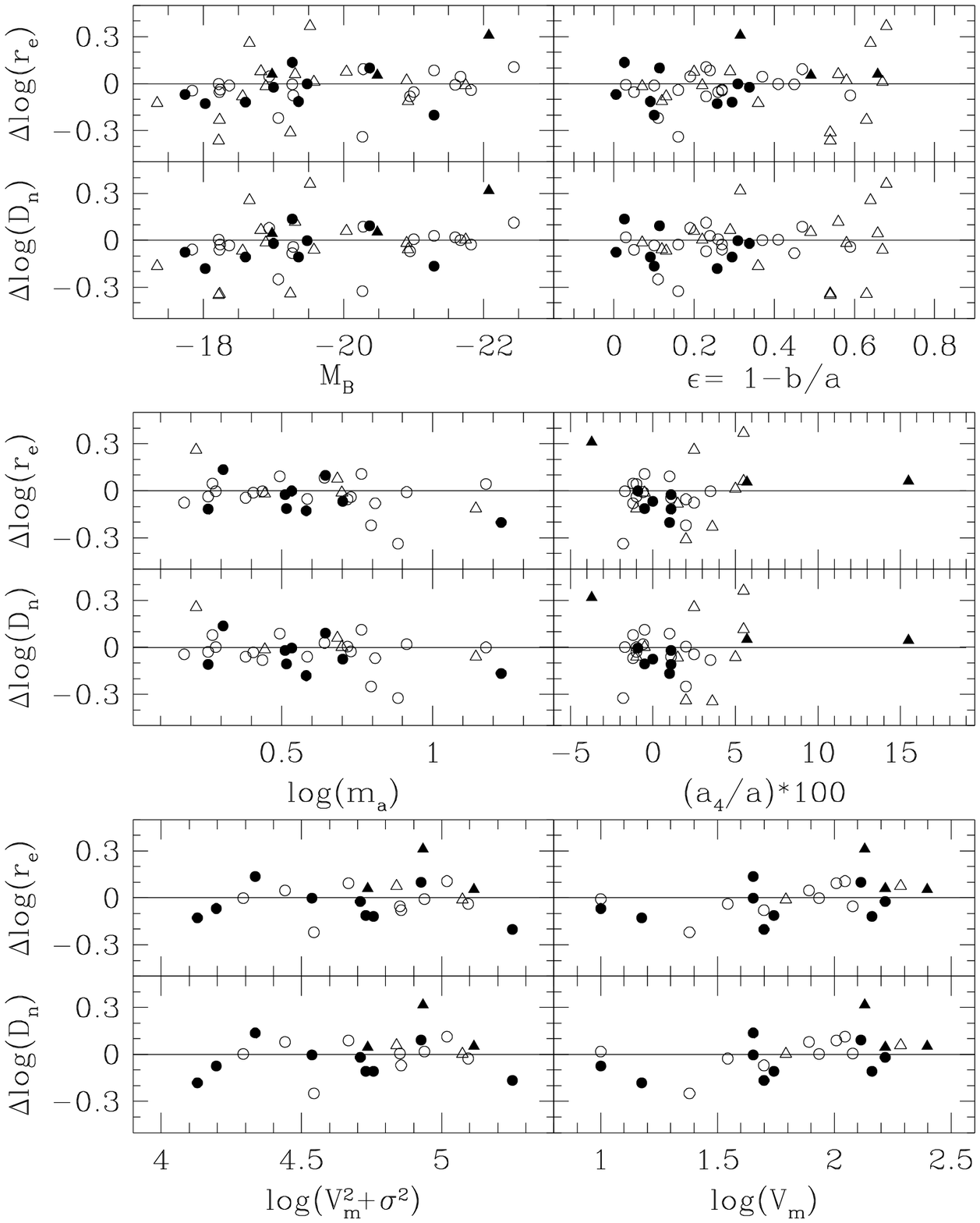,width=17cm,angle=0}
\caption[]{The residuals $\Delta\log(\dn)$ and $\Delta\log(\re)$ of the
\dns\ and FP relations with respect to the absolute magnitude $M_B$, the
ellipticity at \re, $\log(m_{\rm a})$, the boxy/disky parameter
$(a_4/a)*100$, $\log(V_{\rm m}^2+\sigma^2)$, and the maximum observed rotation
velocity along the major axis of the galaxies $\log(V_{\rm m})$. The Virgo and
Fornax galaxies are indicated by open and filled symbols respectively.} 
\label{figresi} 
\end{figure*}

A test of the behaviour of the residuals of the \dns\ and FP relations
is presented in Fig.~\ref{figresi}, where the residuals
$\Delta\log(\dn)$ and $\Delta\log(\re)$ for the \CCD\ galaxies are
plotted versus the galaxy luminosity $M_B$, the ellipticity at the 
effective radius $\varepsilon_{\rm e}$, the normalized coefficient $a_4$ of
the Fourier expansion of the residual of the isophotes with respect to the 
best fitting ellipse (Bender \& M{\"o}llenhoff 1987), the logarithm of the 
exponent $m_{\rm a}$ of the \rn\ law that best fits the major axis light 
profiles of the galaxies, the logarithm of the maximum observed rotation 
velocity along the major axis of the galaxies $V_{\rm m}$, and the logarithm 
of the sum $V_{\rm m}^2+\sigma_0^2$. 

No clear trend of the residuals is seen in almost all cases, with the
possible exception of the galaxies with large values of $m_{\rm a}$ and low
maximum rotation velocities $V_{\rm m}$: galaxies with
$V_{\rm m} < 60$ \kms\ have preferably negative residuals, the opposite being
true for the faster rotators. Unfortunately this result rests on a
small number of objects. The $V_{\rm m}$ values for the Fornax galaxies have
been derived from the rotation curves measured by D'Onofrio \etal\ (1995),
while for Virgo we used the sub-sample of 12 objects with the best
observed rotation curves and velocity dispersion profiles available in
the literature. 

Notably, the small asymmetry in the residuals disappears if we consider
the quantity $V_{\rm m}^2+\sigma^2$, which is more closely related than 
$V_{\rm m}$ to the total kinetic energy.

\begin{figure*} 
\psfig{figure=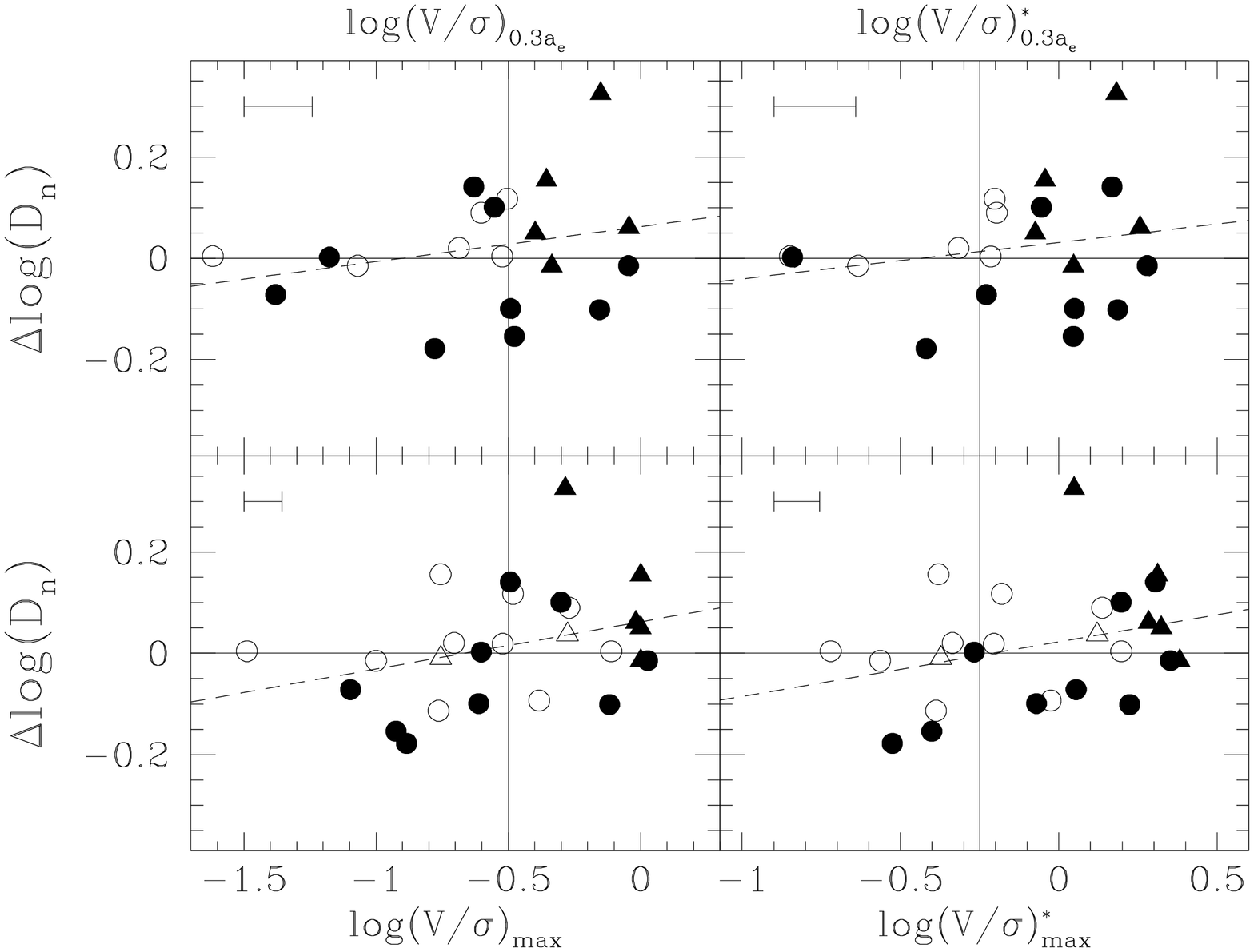,width=17cm,angle=270}
\caption[]{Lower panels: the residuals of the \dns\ relation versus
$\log(V/\sigma)_{\rm max}$ and $\log(V/\sigma)^*_{\rm max}$. Virgo and Fornax
galaxies are represented by open and filled symbols respectively. The
triangles are the S0 galaxies. Upper panels: the residuals versus
$\log(V/\sigma)_{0.3a_{\rm e}}$ and $\log(V/\sigma)^*_{0.3a_{\rm e}}$ for
the same galaxies. The error bars in the upper left corners are averages
of the estimated uncertainties in $V$ and $\sigma$. The dashed lines 
give the best fit to the observed distribution.} 
\label{vsus} 
\end{figure*}

The dependence of the residuals on the kinematical properties of the
galaxies is more clearly seen in Fig.~\ref{vsus}, where $\Delta\log(\dn)$
is plotted versus the ratio $V/\sigma$ between the ordered and random
motions, and versus the anisotropy parameter $(V/\sigma)^*$ (\cf\ Davies
\etal\ 1983), which is a measure of the departure of an object from the
rotational support. In the lower plots $V/\sigma$ is computed using the
highest values assumed by the rotational velocity and the velocity
dispersion in the observed range, while in the upper plots it is the ratio
of the values of $V$ and $\sigma$ measured at the galactocentric distance
$0.3a_{\rm e}$, where $a_{\rm e}$ is the effective semi-major axis of the 
galaxies.

There is possibly a trend in the residuals:
we find $\Dldn = 0.09 \log(V_{\rm m}/\sigma) + 0.06$ with a \rms\ equal to 0.1 
and a correlation coefficient of 0.34.
>From the t-Student distribution the (null) hypothesis of no
correlation between the two variables can be rejected at the 95\% confidence
level. This mild correlation, however, is not completely attributable
to the presence of the S0 galaxies, but rather is due to the different 
distribution of the galaxies dominated by anisotropic pressure 
($\log(V_{\rm m}/\sigma)^*<-0.25$) with respect to those supported by
rotation ($\log(V_{\rm m}/\sigma)^*>-0.25$). 
In fact, if we eliminate the S0 galaxies, a slight trend is still
apparent:  
$\Dldn = 0.06 \log(V_{\rm m}/\sigma) + 0.02$, with a correlation coefficient 
equals to 0.21. 
The trend in the residuals disappears in the correlation with 
$\log(V/\sigma)_{0.3a_{\rm e}}$ when the S0 are eliminated. 
At large radii only the S0 galaxies, that are largely supported by rotation,
have positive residuals. In the upper plot only few Virgo
galaxies have rotation  
curves and velocity dispersion profiles extended to $0.3a_{\rm e}$.

In the above discussion, we did not take into account projection effects in
determining $V_{\rm m}$, since this would require the knowledge of the 
intrinsic structure of each galaxy. This is of course incorrect. However,
Binney (1978, 1980) showed that  
for oblate ellipticals with Hubble luminosity profiles, the ratio 
$\langle V^2 \rangle/\langle \sigma^2 \rangle$ can be approximated to better 
than 20\% by $V^2_{eq}/\sigma^2_c$ (where $V^2_{eq}$ is the true maximum 
equatorial rotation velocity and $\sigma_c$ is the velocity dispersion measured
at the centre of the galaxy), and that $V^2_{eq}/\sigma^2_c = (16/\pi^2)
V^2_0/\sigma_0$, (where $V^2_0$ and $\sigma_0$ are the projected values for
an edge-on oblate galaxy).
Since we can assume a random orientation of the rotation axes with 
respect to the line of sight, the mean projection of $V^2_0$ along this 
line is $V^2_{\rm m} = 2/3 V^2_0$, and we can believe that, to a first order 
approximation, the use of $V_{\rm m}$ and $\sigma_0$ should not produce a 
systematic error of $\langle V^2 \rangle/\langle \sigma^2 \rangle$.

Projection effects along the line of sight are also important. In this case
rotation velocities are generally underestimated (by a factor 1.5-2), while
velocity dispersions remain almost unchanged (Busarello \etal\ 1992).
We have not tried here any deprojection since this would require the
knowledge of the rotation axis of the galaxy with respect to the line of
sight. The measured $V/\sigma$ ratios are therefore lower limits.

Unfortunately the small number of data does not permit a full analysis 
of the effects of rotation on the \dns\ and FP relations. We can only state
that our result is along the same line of that of Saglia \etal\ (1993), who
claim that the presence of disk components inside ellipticals produce a
scatter in the FP, and is in agreement with that of
Prugniel \& Simien (1994), who found the same correlation of
$V/\sigma$ with the residuals of the FP, using a 
sample of galaxies which is not volume-limited.

Indeed, despite the poor correlation, it seems that the kinematical
structure of the 
galaxies plays an important role in the \dns\
and the FP relations, in particular for what concerns the thickness.
The amount of the detected deviations is in agreement with the 
prediction of the Virial theorem, when the contribution of rotation to 
the global equilibrium of elliptical galaxies is taken into account.

\begin{figure} 
\psfig{figure=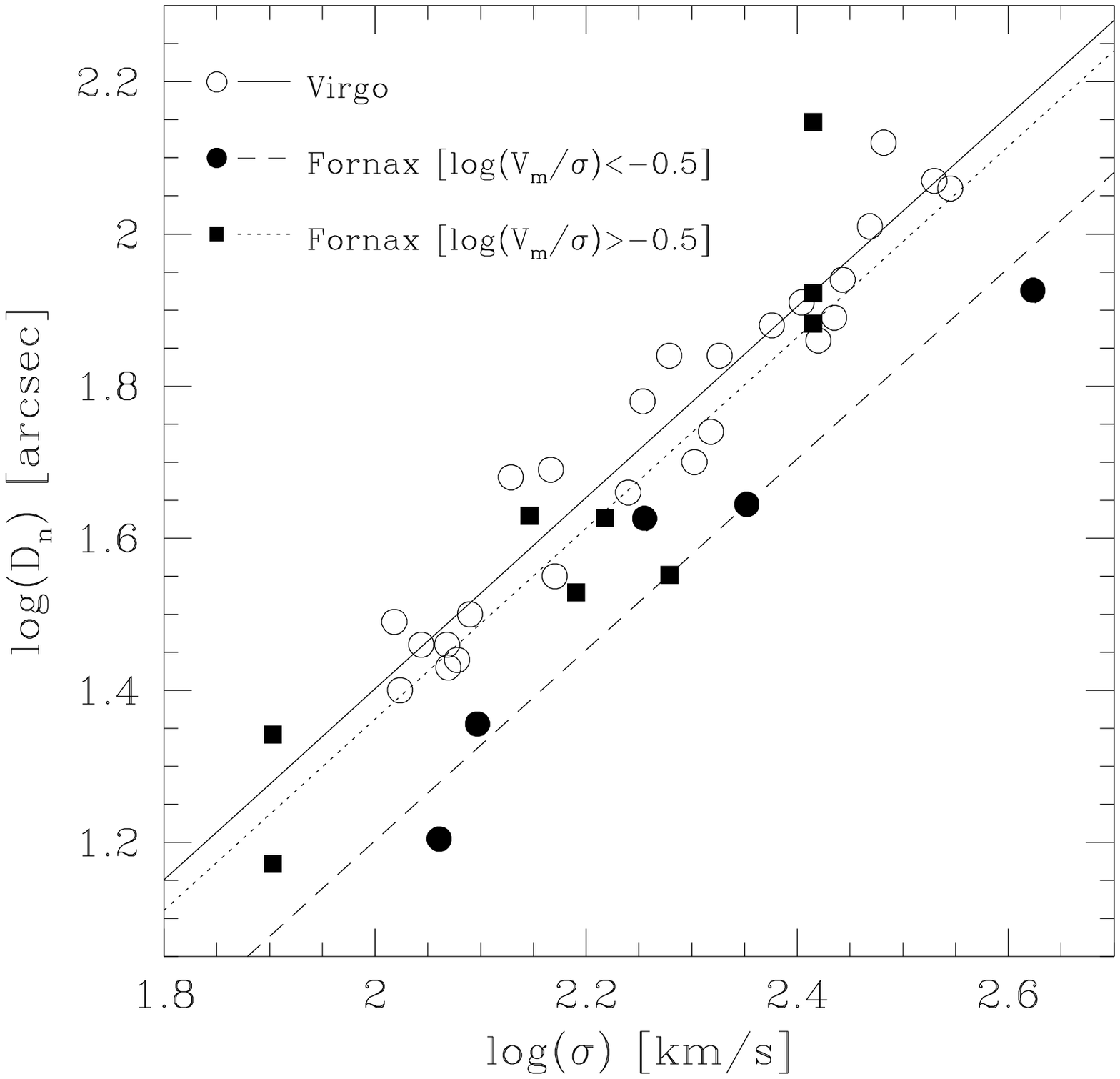,width=8cm,angle=0}
\caption[]{The \dns\ relation for the Virgo and Fornax clusters. The 
galaxies of Fornax are plotted with different symbols according to the 
value of $\log(V/\sigma)_{\rm max}$. Two Fornax galaxies with $\sigma_0 < 100$ 
\kms\ have been added here to increase the number datapoints. 
The larger scatter of the Fornax data with respect to Virgo is more 
evident in this plot.} 
\label{duedn} 
\end{figure}

Fig.~\ref{duedn} shows that the rotationally supported galaxies
provide a mean relative distance modulus for the Fornax cluster with 
respect to 
Virgo which is lower (\Dmufv\ $=$ 0.2 mag) than that obtained from the
galaxies dominated by velocity anisotropy (\Dmufv\ $=$ 1.0 mag). 
Actually, this result is uncertain due to the large scatter of the Fornax
galaxies and the small number of objects, and should be considered here
only as an example of the bias that can be introduced by the rotation effects.
If confirmed by future data, this source of bias should be taken into account 
in the \dns\ and the FP relations when used as DIs of high redshift
clusters.
The misclassification of S0 galaxies would in fact introduce a
systematic bias.

Can we correct the \dns\ for such effect\,? A simple way could be the
following. Let us write that $\dn \propto (\sigma^2)^\alpha$, which means
that the diameter of a galaxy is strictly related to the central velocity
dispersion. If the contribution of the rotation cannot be neglected, we can
write that $\dn' \propto(\sigma^2 + \beta V^2)^{\alpha '}$, where $\beta$
parametrizes the degree of rotational support. Taking the logarithm of these
quantities and assuming that $\alpha ' = \alpha \pm\Delta\alpha$, one
obtains:
\begin{equation}
\label{eq99}
\Delta\log(\dn) = (\alpha\pm\Delta\alpha)\log(1+\beta k^2),
\end{equation} 
where $k=V/\sigma$.

\begin{figure} 
\psfig{figure=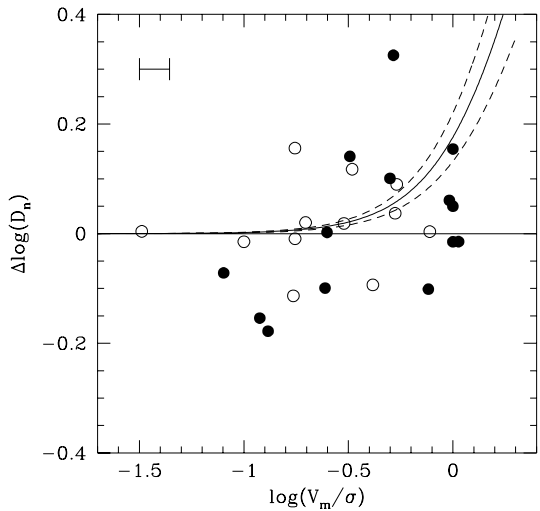,width=8cm,angle=0}
\caption[]{The solid line is the correction curve for the \dn\ diameters
according to eq.~\ref{eq99}, with $\alpha=1$ and $\beta=0.5$. The dashed
lines give the curves for $\Delta\alpha = \pm0.25$ and the same $\beta$.}
\label{corr} 
\end{figure}

Unfortunately we do not know the correct values for the parameters
$\alpha$ and $\beta$. The rotational support, parametrized by $\beta$, is
different for each galaxy, but a mean values for early-type galaxies
could be within 0.5 and 1. The same interval is probably spanned by
$\alpha$ (for the classical \dns\ relation $\alpha\sim 0.6$). Using the
approximation $\alpha=1$ and $\beta=0.5$, one obtains the correction
curves plotted in Fig.~\ref{corr}.

After applying this correction, the trend of the residuals of the \dns\
relation disappears, but the slope of the fitted distribution increases
from 1.26 to 1.37, thus affecting the distance determinations of the
clusters.
Unfortunately the \rms\ of the fitted distribution does not decrease 
significantly.

By considering only the sample of galaxies possessing both $\sigma_0$ and
$V_{\rm m}$, we find that these two variables are not mutally correlated
and that $\log(V_{\rm m}/\sigma)^*$ does not correlate with  
$M_B$, while it correlates with \muem\ (as in Wyse \& Jones 1984). It is also
interesting to note that the residuals of the \dns\ and FP relations,
\Dldn\ and \Dlre, are mutually correlated. We find $\Dlre = 0.94 \Dldn +
0.05$ with an \rms\ of 0.036. This is an obvious consequence of the
Dressler \etal\ relation $\dnae \propto \Iem^{0.8}$. The two
relations are therefore almost equivalent when used as DIs for nearby
clusters.

\subsection{Further test on the \dns\ relation}\label{sec5c}
Do variations in the definition of \dn\ (\ie\ in the surface brightness
level \muen) affect the \dns\,? We performed an experiment with the Virgo
and Fornax \CCD\ sample by varying the value of \dn, \ie\ by assuming that
\muen\ changes from $=18.75$ to 22.75 mag in steps of 1 mag. In the case
of Virgo the slope is found to have small variations, from 1.21 to 1.31,
while for Fornax the variation is larger, and range from 1.19 to 1.37. 
This is due to the larger spread in the central velocity dispersions. The
\rms\ of the fit varies from 0.07 to 0.11 for Virgo and from 0.13 to 0.16
for Fornax. The best fits occur at $\mu_B = 19.75$ and 20.75 \mga. All
these expressions for the \dns\ relation provide approximately the same
$\Delta\log(\dn)$ (it varies from 0.09 to 0.11), that is the same \Dmu.

In another experiment we have acted on the definition of $\sigma$,
adopting once the value of the velocity dispersion measured at the centre,
$\sigma_0$, and another time the average within a given radial range.
The motivation for this experiment comes from the Virial Theorem,
since we want to look for the value of $\sigma$ that best represents
$\langle \sigma \rangle$.

We decided to average $\sigma$ within circles of radius $r_n=D_n/2$ and 
$0.3 a_{\rm e}$, where $a_{\rm e}$ is the effective semi-major axis.
In each case we took plain averages, weighted means for the surface
brightness, and means weighted for the rotational contribution.
Unfortunately, this test was possible only for the 15 galaxies of the Fornax 
cluster which have accurate extended rotation curves, velocity dispersion
profiles, and luminosity profiles. 
We found that the slopes of the \dns\ relation obtained in each
case vary in the interval 1.34--1.62 (to be compared
with the slope of 1.26 achieved using the central velocity dispersion),
that is it changes by an amount larger that the error of the fit, and
increases with the radius out to which the average is computed. 
Approximately equal \rms\ fits are obtained using the central velocity 
dispersions $\sigma_0$ and the value averaged within 0.3$a_{\rm e}$. 
Slightly larger errors are derived with the means within $r_n$.

These results are an indication that early-type galaxies deviate from the
pure homology and that the \dns\ and the FP relations have a slope and an
intrinsic scatter connected to the spatial and dynamical structure of the
galaxies.

One might try to use the quantity $(V_{\rm m}^2+\sigma^2)$ instead of $\sigma$
in order to account for the above mentioned bias introduced by the
different kinematical properties of the galaxies. The result of such an
experiment is that the slope of the relation changes from $\sim1.23$ to
$\sim1.37$, but the \rms\ of the fit does not show a significant
improvement.

\section{The \lrem\ relation}\label{sec6}
\begin{figure} 
\psfig{figure=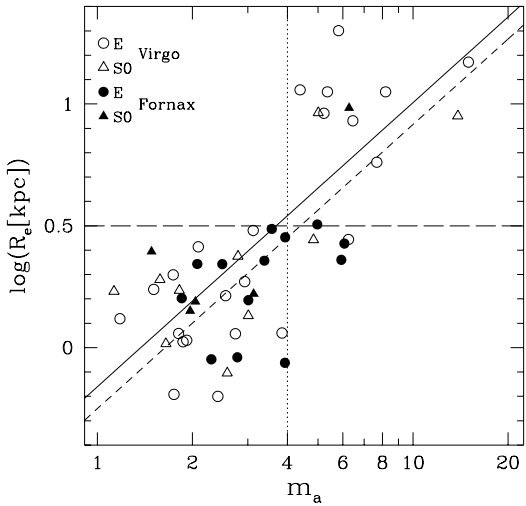,width=8cm,angle=0}
\caption[]{The \lrem\ relation for the Virgo and Fornax galaxies. 
$R_{\rm e}$ is the effective radius of the galaxies in kpc. The solid 
line gives the best fit for the Virgo galaxies. The dashed line is the fit
of minimum scatter for the Fornax sample.}
\label{lognlogre}
\end{figure}

We finally present another method to derive the relative distance to the
Virgo and Fornax clusters. This method is based on the \lrem\ relation
(Caon \etal\ 1993, Young \& Currie 1994) and is shown in
Fig.\ref{lognlogre}. Here $m_{\rm a}$ is the exponent of the \rn\ Sersic'
law best 
fitting the major axis surface brightness distribution of the galaxies,
and $R_{\rm e}$ is the effective radius in kpc. Although the scatter in
this relation is quite large (the \rms\ of the fit is $=0.28$ and the 
correlation coefficient is 0.7), the intercept offset $\Delta\log R_{\rm e}$ 
turns out to be $0.089$, corresponding to $\Dmufv = 0.44$ mag, in excellent 
agreement with the \dns\ and FP relations. 
Such agreement is remarkable although probably fortuitus given the large
scatter of the data in this relation.

\section{Conclusion}\label{sec7}
The numerical results obtained in this work fitting the \dns\ and FP relations 
are summarized in Table~4.
>From these data we derive $\Dmufv=0.45$ mag and $\Dmucv=3.45$ mag respectively
for the relative distance moduli to Fornax and Coma with respect to the
Virgo cluster. 
Taking into account the corrections for the aperture effects \Dmufv\ is
closer to 0.35 mag and \Dmucv\ to 3.55 mag.
Our distance to Fornax is slightly larger than derived with
the other DIs, but it is nearly equal to that derived with the BCM method 
(see Table~1).

\begin{table*}
\label{summary}
\begin{tabular}{lrrrcccc}
\multicolumn{8}{c}{Table 4} \\
\hline\hline \\[-7pt]
Method & {a\ \ \ \ \ \ \ } & {b\ \ \ \ \ \ \ }& {c\ \ \ \ \ \ \ } & rms & 
$\Dmufv$ & $\Dmucv$ & Notes \\
\hline \\[-7pt]
\dns\ & $1.26\pm0.06$ & $-1.11\pm0.14$ & & 0.06 & 0.55 &      & \ph{1}1 \\
\dns\ & $1.31\pm0.07$ & $-1.24\pm0.16$ & & 0.06 & 0.45 & 3.40 & \ph{1}2 \\
\dns\ & $1.24\pm0.06$ & $-1.08\pm0.15$ & & 0.08 & 0.45 & 3.45 & \ph{1}3 \\
\dns\ & $1.26\pm0.06$ & $-1.11\pm0.14$ & & 0.06 & 0.46 & 3.46 & \ph{1}4 \\
\dns\ & $1.37\pm0.31$ & $-1.46\pm0.69$ & & 0.13 & 0.47 &      & \ph{1}5 \\
\dns\ & $1.16\pm0.08$ & $-1.59\pm0.19$ & & 0.08 &      & 3.45 & \ph{1}6 \\
\lsmu\ & $1.35\pm0.08$ & $-0.008\pm0.178$ & $0.002\pm0.016$ & 0.08 & 0.45 &
3.45 & \ph{1}7 \\
FP & $1.26\pm0.09$ & $0.28\pm0.01$ &$-7.31\pm0.28$& 0.06 & 0.45 &3.45&\ph{1}8 \\
FP & $1.01\pm0.03$ & $0.07\pm0.04$ & & 0.06 & 0.30 & 3.45 & \ph{1}9 \\
FP & $0.85\pm0.10$ & $0.23\pm0.13$ & & 0.13 & 0.35 &      & 10 \\
FP & $1.15\pm0.04$ &$-0.84\pm0.06$ & & 0.06 &      & 3.50 & 11 \\ 
\lrem\ & $1.16\pm0.25$ & $-0.160\pm0.14$ & & 0.28 & 0.44 & & 12 \\
\hline \\
\end{tabular}

\footnotesize{
Notes: 1) Linear least square fit for the Virgo galaxies and fit of
minimum scatter for Fornax adopting the same slope of Virgo.
2) Fit to the Virgo galaxies taking into account the errors on both 
coordinates.
3) Fit of minimum scatter for Fornax and Coma using the fit of the combined
sample of 128 galaxies.
4) Output of program CALIB using Virgo as calibrator.
5) Output of program CALIB using Fornax as calibrator.
6) Output of program CALIB using Coma as calibrator.
7) Coefficients of the \lsmu\ relation from the multivariate statistical
analysis. Fit of minimum scatter for Fornax and Coma, adopting the same slope.
8) Coefficients of the FP relation from the multivariate statistical 
analysis. Fit of minimum scatter for Fornax and Coma, adopting the same slope.
9) Output of program CALIB using Virgo as calibrator.
10) Output of program CALIB using Fornax as calibrator.
11) Output of program CALIB using Coma as calibrator.
12) Coefficients of the fit for the Virgo galaxies. Fit of minimum scatter
for Fornax, adopting the same slope.}
\end{table*}

The inner ring diameters of spirals, the globular cluster luminosity
function (GCLF), the surface brightness fluctuations (SBF), the \lsmu\
relation, the IR Tully--Fisher relation (\IRTF), and the scale length of
dE galaxies, are in open contrast with this determination, while the
planetary nebulae luminosity function (PNLF), and the SNe--Ia are consistent.

For the Coma cluster we derived a distance smaller than given by the other
DIs, but now a better agreement exists with the TF relation.

Probably our estimates can be improved when new accurate and homogeneous
central velocity 
dispersions and rotation velocities will be available for a complete sample of
Virgo and Fornax early--type galaxies.
Actually, the large intrinsic scatter we found for the \dns\ and FP
relation of Fornax is at variance with previous determinations.
We believe that the problem is in the measurements of $\sigma$:
the large spread of our central velocity dispersions being caused by
projection effects, $\sigma$ anomalies, and the presence of inner 
sub-components such as disks or central black holes.
In this context, the large uncertainty on \Dmu\
resides probably in the different procedures
used for measuring the velocity dispersions and in the corrections adopted
for taking into account the effects of the aperture sizes.

By applying an average aperture correction to the central velocity dispersions,
we obtain differential distance moduli in better agreement with the
literature data, and we find that the tilt of the \dns\ relation increases.

Looking at the data presented in this paper we conclude that the \dns\ and
the FP relation give substantially the same result. An advantage of the
\dns\ relation over the FP is that \dn\ can be measured with high
photometric accuracy, while the effective radius \re\ and 
\muem\ may depend on the actual shape of the luminosity
profiles and on the error in the total luminosity of the galaxies. 
The errors in \re\ and \muem\ are in fact usually correlated (Capaccioli
\etal\ 1992).

At variance with van Albada \etal\ (1993), we do not find any need for
adding a surface brightness correction to the \dns\ relation, at least for the
galaxies of the \CCD\ sample.
The inverse \dnaemue\ relation of Dressler \etal\ (1987),
coupled with the observed \dns\ and FP relations, 
has indicated that the residuals of the \dns\ relation depend only slightly 
on \muem, since all the coefficients in eq.~\ref{eq9} are approximately null.

The residuals in both the \dns\ and FP relations are found to be
independent of the intrinsic luminosity of the galaxies, the ellipticity,
the mean surface brightness and the boxiness/diskiness parameter \aq. 
A trend of the residuals, on the other
hand, is present with the exponent $m$ of the \rn\ Sersic's law, with the
maximum rotation velocity of the galaxies and with 
the $V_{\rm m}/\sigma$ ratio. The last result is in agreement with
Busarello \etal\ (1992) and Prugniel \&
Simien (1994), who first pointed out the contribution of rotation to the
global equilibrium of elliptical galaxies.

The correlation of the residuals of the \dns\ relation with the $(V/\sigma)$
ratio, with the anisotropy parameter $(V/\sigma)^*$, and with the exponent
$m_{\rm a}$ of the major axis light profiles of the galaxies are all
indications of the non-homologous nature of the early-type galaxies.

Further improvements call for more accurate kinematical measurements, for
a critical estimate of the different methods of data reductions, for a
standard correction of the aperture effects, and
for a deeper analysis of the kinematics of each individual galaxy. 
Even at this 
stage, we are probably legitimated to say that most of the observed
deviations from the FP are likely due to departures from the pure homology in
the density distribution and in the dynamical structure of the early-type
galaxies.

An important step for the use of the \dns\ and the FP relations as DIs will
be that of defining a ``representative sample'' of galaxies, of using standard
selection criteria, and data with the same degree of homogeneity and
accuracy, as well as variables properly and univocally determined.

In closing we want to remark that, while our analysis disagrees with
Saglia's \etal\ (1993) for what concerns the dependence of the residuals
on the ellipticity and on $a_4$, nonetheless we back their conclusion: the
kinematical structure of the early--type galaxies, if not properly taken
into account, may produce biased cluster distances, when the \dns\ and FP
relations are used as DIs without care.

\end{document}